\documentclass{iopjournal}
\usepackage{academicons}
\usepackage{amsmath,mathrsfs}
\usepackage{eufrak,amsfonts,txfonts}
\usepackage{ragged2e}
\usepackage[final]{microtype}
\newcommand{\doi}[1]{(\href{https://doi.org/#1}{doi: \nolinkurl{#1}})}
\newcommand{\orcidlink}[1]{\href{https://orcid.org/#1}{\includegraphics[width=8pt]{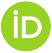}}}
\hypersetup{colorlinks=true,linkcolor=blue,citecolor=blue,urlcolor=blue}
\newtheorem{theorem}{Theorem}[section]
\newtheorem{definition}{Definition}[section]
\newtheorem{corollary}{Corollary}[theorem]
\newtheorem{lemma}[theorem]{Lemma}
\newtheorem{proposition}[theorem]{Proposition}
\newtheorem{remark}[theorem]{Remark}

\begin{document}
\fancyhead[R]{Fülöp Bazsó}

\articletype{Paper}
\title{Zero-divisor algebras of graph functions: quantum caging, entangling routing and stochastic first-passage exclusion}

\author{Fülöp Bazsó \orcid{0009-0002-0931-6266}}

\affil{Department of  Computational Sciences\\ HUN-REN Wigner Research Centre for Physics, Institute for Particle and Nuclear Physics \\
        P.O. Box 49, H-1525, Budapest, Hungary}
	\email{bazso.fulop@wigner.hu}



\keywords{graph Lie algebras, zero divisors, Lie bialgebras, quantum walks, first-passage processes, algebraic caging, graph decorations, stochastic routing}

\begin{abstract}
We show how algebraic constraints on weighted graphs yield exact localization and controlled transport in quantum and stochastic systems. Antisymmetrizing a noncommutative, nonassociative product of vertex functions defines a Lie bracket. For a specified Lie cobracket over commutative rings without nonzero $2$-torsion, compatibility holds if and only if products of distinct incident edge weights vanish. When the coefficient ring is a finite product of integral domains, the edges with nonzero weights in each component of the ring form a matching. These matchings are determined by the edge weights and need not coincide across the components.

These matchings yield exact two-vertex caging in fixed quantum channels; separately specified channel-selective controls enable coherent routing and path--internal-state entanglement. For matching-supported stochastic jump rates, they enforce first-passage exclusion, while prescribed internal-state switching opens inaccessible paths and yields a singular weak-switching escape limit. The bracket--cobracket factorization identifies a common edge operator underlying quantum kinetic Hamiltonians and Markov generators, subject to distinct sign and positivity conditions. The algebra fixes the channel structure; controlled motion requires additional dynamical operations.
\end{abstract}


\section{Introduction}

Difference and Laplace operators on finite graphs are central to discrete geometry, stochastic processes and quantum transport \cite{ch,sun}. A graph difference, however, is not a derivation of the pointwise product of vertex functions. This motivates the search for an algebraic product that generates graph operators intrinsically while satisfying an exact derivation identity. Related approaches include discrete differential calculi, noncommutative geometry and metric quantum graphs \cite{dim,majid,berkuch,rovelli2019}. Our setting is purely combinatorial: functions and wavefunctions are defined on the vertices of a finite graph.

We consider the antisymmetric product
\begin{equation}
 [f,g]_i=\sum_j w_{ij}(f_i g_j-f_j g_i)
\end{equation}
for which the weighted graph Laplacian becomes the inner derivation $\operatorname{ad}_1=-\Delta_w$. For this product to define a Lie bracket, its edge weights must satisfy Jacobi's identity. Over an integral domain, the restriction is strong: whenever two edges with nonzero weights share a vertex, their other endpoints are also joined by an edge and all three weights agree. Consequently, the support of nonzero weights is a disjoint union of uniformly weighted cliques. Zero divisors provide a way beyond this rigidity. A proper edge colouring assigns different colours to any two distinct edges incident at the same vertex. If properly coloured edges carry mutually orthogonal idempotents, products of weights on distinct incident edges vanish while squares of individual weights remain nonzero. Vizing's theorem \cite{diestel} guarantees that every finite simple graph admits a proper edge colouring using at most one more colour than its maximum vertex degree. Assigning mutually orthogonal idempotents to these colours yields the required edge weights.

The paper makes four connected contributions. First, it establishes necessary and sufficient conditions for the graph bracket to satisfy Jacobi's identity and characterizes the rigidity and realizability of the resulting Lie algebras. Second, it establishes necessary and sufficient conditions for compatibility with the specified canonical edge-difference cobracket. Compatibility holds precisely when products of weights on distinct incident edges vanish, provided the coefficient ring has no nonzero $2$-torsion. For finite products of integral domains, this has a particularly simple consequence: the nonzero edge weights in each coefficient component must be supported on a matching. In the reduced idempotent realization, the resulting Lie algebra decomposes into two-dimensional affine edge factors and central isolated directions. It is metabelian and, whenever a nonzero edge factor occurs, nonnilpotent.

Third, direct proofs of the co-Jacobi identity and the $1$-cocycle identity lead to the factorizations $\Delta_w=\partial\delta$ and $\Delta_w^2=2\delta^\dagger\delta$. The resulting positive edge operator $Q=\delta^\dagger\delta$ generates the quantum kinetic Hamiltonian after suitable scaling and, with the opposite sign and prescribed nonnegative rates, the pure-diffusion Markov generator. These are distinct physical operators, despite their common algebraic origin. Fourth, the matching restriction has complementary dynamical realizations. In a fixed quantum channel it yields exact two-site caging; independently supplied channel-selective controls permit coherent transfer and path--internal-state entanglement. For stochastic dynamics, matching-supported transition rates give first-passage exclusion and channel-dependent absorbing routing. Separately prescribed channel switching breaks that confinement and leads to a singular crossover to metastable escape. These applications connect with graph state transfer \cite{bose,christandl} and first-passage and switching-diffusion theory \cite{redner,benichou,grebenkov2019}. Unlike Aharonov--Bohm cages \cite{vidal}, the localization here follows from the algebraic channel restriction rather than flux-tuned interference.

The matching realization works on arbitrary finite graphs and on iterated decorations of bounded degree. Appendix~\ref{app:dark} develops complementary spectral examples, including a heterogeneous cycle-chain construction and an analysis of converter conditions on nonplanar $K_{3,3}$, under explicit assumptions on the active channels, converters and observation ports. Channel conversion and cage-preserving interactions are independent physical controls that extend the dynamics supported by the algebraic construction.

We begin with the Lie algebra and its compatible cobracket, then develop the quantum and stochastic applications. In the quantum realization, orthogonal internal-state projectors represent the coefficient idempotents on a conventional Hilbert space. The stochastic realization instead uses a joint process on position and a classical internal label. In both settings the coefficient algebra has a concrete channel interpretation, providing a common formalism for distinct, explicitly defined dynamics.

\subsection{Preliminaries}

Let $G=(V,E)$ be a finite simple graph and let $R$ be a commutative unital coefficient ring. The degree $\deg(v)$ of a vertex $v$ is the number of edges incident to it, and $\Delta(G)=\max_{v\in V}\deg(v)$ denotes the maximum vertex degree. The vertex module is
\begin{equation}
 \mathcal H_V=R^V,
\end{equation}
with standard basis $e_i(j)=\delta_{ij}$.  Assign symmetric edge weights
\begin{equation}
 w_{ij}=w_{ji}\in R,\qquad w_{ii}=0,
\end{equation}
with $w_{ij}=0$ when $ij\notin E$.  The corresponding weighted graph Laplacian is
\begin{equation}
 (\Delta_w f)_i=\sum_jw_{ij}(f_i-f_j).
 \label{eq:weighted_graph_laplacian}
\end{equation}
For ordinary nonnegative real weights, this sign convention makes $\Delta_w$ positive semidefinite.  The constant function $1=\sum_i e_i$ lies in its kernel.

In fact the pointwise Leibniz defect is
\begin{equation}
 (\Delta_w(fg))_i-f_i(\Delta_w g)_i-g_i(\Delta_w f)_i
 =-\sum_jw_{ij}(f_i-f_j)(g_i-g_j),
 \label{eq:pointwise_leibniz_defect}
\end{equation}
which is generally nonzero.  The Laplacian is an inner derivation of the Lie bracket, not of the pointwise function product.

The pointwise product on $R^V$ does not make $\Delta_w$ a derivation.  The aim is instead to construct an antisymmetric product for which $\Delta_w$ is generated internally and obeys the exact derivation identity associated with a Lie bracket.  The algebraic results below hold over a general commutative ring.  For the physical applications we specialize to $\mathbb C^q$ and $\mathbb R^q$, whose orthogonal idempotents provide non-nilpotent zero divisors and componentwise positivity.

\section{Lie algebras}

The present construction shares with the earlier Boolean--Lie algebra of Bazs\'o and L\'abos~\cite{bla} the idea of obtaining a Lie bracket by antisymmetrizing a noncommutative, nonassociative product, although the underlying products and the mathematical questions investigated are different. Kauffman's noncommutative-world formalism~\cite{kauf} provides a different operator-commutator approach to discrete derivatives, not a classification of this vertex bracket; Novikov's graph Schr\"odinger operators and symplectic geometry~\cite{nov} address spectral geometry, not the question of whether this product satisfies Jacobi's identity. For the specific bracket below, the contributions are the exact Jacobi test, the classification of clique supports over products of integral domains and the necessary-and-sufficient compatibility criterion for the specified same-weight edge-difference cobracket. The bracket is

\begin{definition}
  \begin{eqnarray}
  && [\; \; , \; ] : \mathcal{H}_V \times \mathcal{H}_V \rightarrow \mathcal{H}_V \nonumber \\
  && [f,g]_i = \sum_j w_{ij} (f_i g_j - f_j g_i) \label{eq:commutator}
\end{eqnarray}
\end{definition}

The bracket is bilinear and antisymmetric.

The bracket can be rewritten in terms of the Laplace operator:
\begin{eqnarray}
  [f,g]_i = (\Delta_w f)_i g_i-f_i(\Delta_w g)_i \label{eq:Comm_Laplace}
\end{eqnarray}
One has $\operatorname{ad}_1(f)=[1,f]=-\Delta_wf$ for the constant function $1$.  Thus the weighted graph Laplacian is an inner derivation of the Lie algebra.

The commutator of base functions is:
\begin{eqnarray}
  [e_a,e_b]   = w_{ab} e_a - w_{ab} e_b \label{eq:e_com_rel}
\end{eqnarray}

The matching structure used in the algebraic construction is provided by proper edge colouring.  Let $R$ be a commutative unital ring and set $w_{ij}=w_{ji}$ and $w_{ii}=0$.  The condition that products of weights on distinct incident edges vanish is
\begin{equation}
 w_{ij}w_{ik}=0\qquad (j\neq k),
 \label{eq:incident_annihilation}
\end{equation}
for every pair of distinct edges incident at the same vertex.

\begin{theorem}[Criterion for Jacobi's identity]\label{thm:jacobi_exact}
The bracket in eq.~(\ref{eq:commutator}) satisfies Jacobi's identity if and only if, for every three distinct vertices $a,b,c$,
\begin{equation}
\begin{aligned}
 w_{bc}(w_{ab}-w_{ac})&=0,\\
 w_{ac}(w_{bc}-w_{ab})&=0,\\
 w_{ab}(w_{ac}-w_{bc})&=0.
\end{aligned}
\label{eq:jacobi_weight_conditions}
\end{equation}
In particular, the incident-annihilation condition (\ref{eq:incident_annihilation}) is sufficient.
\end{theorem}
\textbf{Proof}: From eq.~(\ref{eq:e_com_rel}), direct expansion on three distinct basis elements gives
\begin{equation}
\begin{split}
 \operatorname{Jac}(e_a,e_b,e_c)
={}&w_{bc}(w_{ab}-w_{ac})e_a
+w_{ac}(w_{bc}-w_{ab})e_b\\
&+w_{ab}(w_{ac}-w_{bc})e_c.
\end{split}
\end{equation}
Linear independence of the basis yields eqs.~(\ref{eq:jacobi_weight_conditions}).  Under eq.~(\ref{eq:incident_annihilation}), each coefficient is a difference of two products of distinct incident edge weights and therefore vanishes.  Cases with repeated vertices cancel by antisymmetry. \hfill Q.E.D.

\begin{corollary}[Rigidity over integral domains]\label{cor:domain_rigidity}
If $R$ is an integral domain and every edge in the support has nonzero weight, then every connected component of the support graph is either an isolated vertex, a single edge, or a complete graph whose edges all carry the same weight.
\end{corollary}
\textbf{Proof}: If $a-b-c$ is a path of nonzero support edges, then $w_{ab}\neq0$ and $w_{bc}\neq0$.  The first equation in (\ref{eq:jacobi_weight_conditions}) gives $w_{ab}=w_{ac}$ by cancellation of the nonzero factor $w_{bc}$, and hence $w_{ac}\neq0$.  The third equation then gives $w_{ac}=w_{bc}$ by cancellation of $w_{ab}$.  Thus all three edge weights are equal and nonzero.  Repeating the argument along paths shows that every connected component is complete and uniformly weighted. \hfill Q.E.D.

\begin{corollary}[Universal zero-divisor realization]\label{cor:universal_colouring}
Let $c:E(G)\rightarrow\{1,\ldots,q\}$ be a proper edge colouring, and write $c(i,j):=c(\{i,j\})$ for the colour of the undirected edge $\{i,j\}$. Let
$R=\mathbb C^q=\bigoplus_{\alpha=1}^q\mathbb C\varepsilon_\alpha$, where
$\varepsilon_\alpha\varepsilon_\beta=\delta_{\alpha\beta}\varepsilon_\alpha$.
We call the component labelled by $\alpha$ a channel. The edges whose weights have a nonzero $\varepsilon_\alpha$ component determine the corresponding channel subgraph; under the compatibility condition for the specified cobracket, this subgraph is a matching. In the quantum and stochastic realizations, channels are represented by internal particle states. For arbitrary nonzero scalars $\gamma_{ij}$, the weights
\begin{equation}
 w_{ij}=\gamma_{ij}\varepsilon_{c(i,j)}
 \label{eq:idempotent_edge_weights}
\end{equation}
satisfy Jacobi's identity.  Consequently every finite simple graph admits such a realization with $q\leq\Delta(G)+1$.
\end{corollary}
\textbf{Proof}: Distinct incident edges have different colours, so their weights multiply to zero.  The conclusion follows from Theorem~\ref{thm:jacobi_exact} and Vizing's theorem. \hfill Q.E.D.

\begin{corollary}[Classification over products of integral domains]\label{cor:coordinatewise_cliques}
Let $R=\prod_{\alpha=1}^q R_\alpha$ be a finite product of commutative integral domains, and let the symmetric edge weights $w_e\in R$ be arbitrary (some components may vanish).  The bracket satisfies Jacobi's identity if and only if, for every $\alpha$, the support of the scalar weights $w_e^{(\alpha)}$ is a disjoint union of uniformly weighted complete graphs and isolated vertices.  The complete subgraphs determined by the nonzero edge weights in different components of the coefficient ring may share vertices and edges.  For $R=\mathbb C^q$ this classifies all Jacobi-compatible weights, not merely those arising from proper edge colourings.
\end{corollary}
\textbf{Proof.} Since $R$ is a finite product of integral domains, the three equations expressing Jacobi's identity hold in $R$ if and only if they hold in each component of the coefficient ring. Corollary~\ref{cor:domain_rigidity} then gives the stated structure for the subgraph selected by the nonzero edge weights in each component. Conversely, uniformly weighted complete graphs satisfy these equations. Hence the equations hold in every component, and therefore in $R$. \hfill Q.E.D.

As an example of overlapping clique components, take the diamond with vertices $a,b,c,d$ and edges $ab,ac,bc,ad,bd$.  Over $\mathbb C^2$ assign
\begin{equation}
 w_{ab}=\varepsilon_1+\varepsilon_2,\qquad
 w_{ac}=w_{bc}=\varepsilon_1,\qquad
 w_{ad}=w_{bd}=\varepsilon_2.
\end{equation}
The first component of the coefficient ring supports the uniformly weighted triangle $abc$, and the second supports the triangle $abd$, so Jacobi's identity holds.  Nevertheless $w_{ab}w_{ac}=\varepsilon_1\ne0$, showing that Jacobi's identity alone does not impose the matching condition required by the specified cobracket. This example illustrates how the subgraphs selected by different components of the coefficient ring can overlap.

Jacobi's identity permits uniformly weighted clique components. Compatibility with the specified cobracket requires that, in each component of the coefficient ring, no two edges with nonzero weights share a vertex. In particular, the complete subgraphs selected by different components may overlap without violating Jacobi's identity. Orthogonal idempotents provide one useful construction, but do not exhaust the solutions of Jacobi's identity.

The graph determines which pairs of vertices contribute to the bracket, while the edge weights determine the corresponding contributions in the coefficient ring. In the realization (\ref{eq:idempotent_edge_weights}), edges assigned to different channels have weights whose products vanish.

From eq.~(\ref{eq:e_com_rel}) the only nonzero structure constants are
\begin{equation}
 f_{ab}^{a}=w_{ab},\qquad f_{ab}^{b}=-w_{ab}.
\end{equation}
For each channel $\alpha$, let $\mathcal M_\alpha$ be the support of its weights and let $I_\alpha$ be the set of vertices not covered by that matching.

\begin{theorem}[Channel decomposition]\label{thm:channel_decomposition}
For arbitrary canonically compatible weights $w_e=\sum_{\alpha=1}^q\gamma_{e,\alpha}\varepsilon_\alpha$ over $\mathbb C^q$, define $\mathcal M_\alpha=\{e:\gamma_{e,\alpha}\ne0\}$. By the canonical compatibility criterion of Theorem~\ref{thm:cocycle} (equivalently, Corollary~\ref{cor:complete_matching_layers} below), each $\mathcal M_\alpha$ is a matching (an edge may belong to several $\mathcal M_\alpha$). Here $\mathfrak{aff}(1)$ denotes the two-dimensional nonabelian complex Lie algebra with a basis $h,x$ satisfying $[h,x]=x$; its subscript $ij$ labels the corresponding edge factor.  The notation $\mathbb C^{I_\alpha}$ denotes the abelian Lie algebra spanned by the vertices in $I_\alpha$.  Restriction of scalars to $\mathbb C$ gives
\begin{equation}
 \mathfrak g
 \cong
 \bigoplus_{\alpha=1}^{q}
 \left(
 \bigoplus_{\{i,j\}\in \mathcal M_\alpha}\mathfrak{aff}(1)_{ij}
 \oplus
 \mathbb C^{I_\alpha}
 \right).
 \label{eq:affine_decomposition}
\end{equation}
Hence $[\mathfrak g,\mathfrak g]$ is abelian.  The algebra is metabelian, and it is nonnilpotent whenever at least one edge weight is nonzero.
\end{theorem}
\textbf{Proof}: By Theorem~\ref{thm:cocycle}, and as made explicit in Corollary~\ref{cor:complete_matching_layers}, the support in each channel is a matching. Thus distinct edges in one channel have disjoint vertex supports and commute.  On one nonzero edge $\{i,j\}$, the change of basis $x=e_i-e_j$, $y=e_i+e_j$ gives $[y,x]=-2\gamma_{e,\alpha}x$ in the $\varepsilon_\alpha$ sector, which is the two-dimensional affine Lie algebra.  Channel-isolated vertices are central.  The derived algebra is therefore the direct sum of the one-dimensional spans of the edge differences and is abelian.  Since $\operatorname{ad}_y^n(x)=(-2\gamma_{e,\alpha})^n x$ on a nonzero edge factor, the lower central series does not terminate. \hfill Q.E.D.

\begin{corollary}\label{cor:killing_singular}
The Killing form is well defined but singular.  In the idempotent-valued representation its coefficients are supported in the corresponding channel ideals, and after restriction to $\mathbb C$ each $\mathfrak{aff}(1)$ factor and each central factor contributes a degenerate block.
\end{corollary}

The centre is also read off channel by channel:
\begin{equation}
 Z(\mathfrak g)=
 \bigoplus_{\alpha=1}^{q}
 \operatorname{span}_{\mathbb C}
 \{\varepsilon_\alpha e_i:\ i\in I_\alpha\}.
 \label{eq:center_channel_formula}
\end{equation}
In particular, the centre is trivial if and only if $\mathcal M_\alpha$ is a perfect matching on the full vertex set for every component $\alpha=1,\ldots,q$. An unused component ($\mathcal M_\alpha=\varnothing$) contributes the entire central summand $\varepsilon_\alpha\mathbb C^V$.

\subsection{Weight-preserving symmetries}
A vertex permutation induces a bracket automorphism when it preserves the complete $R$-valued edge weights, including their idempotent components. Thus colouring can break uncoloured graph automorphisms. These weight-preserving automorphisms commute with $\Delta_w$.

\section{Lie bialgebras}

We use ``Lie bialgebra over $R$'' in the finite-free module sense: the bracket and cobracket are $R$-linear and satisfy Jacobi's identity, the co-Jacobi identity and the $1$-cocycle condition.  Let $\mathfrak g$ be the graph Lie algebra over a commutative ring $R$ satisfying Jacobi's identity; the following theorem gives necessary and sufficient conditions for compatibility with the specified cobracket. See \cite{yks,farinati} for general Lie-bialgebra background and related graph constructions.  Here self-dual means that, after identifying the vertex basis with its dual, the bracket and cobracket use the same edge-weight coefficients. The cobracket is called canonical relative to this choice and the weighted graph-difference operator; other cobrackets, including the zero cobracket, are possible. We define an antisymmetric cobracket
\begin{equation}
 \delta:\mathfrak g\longrightarrow\bigwedge^2\mathfrak g,
 \qquad
 \delta(e_i)=\sum_j w_{ij}e_i\wedge e_j.
 \label{eq:cobracket_basis}
\end{equation}
For a general vertex function $h=\sum_i h_i e_i$, this is equivalently
\begin{equation}
 \delta(h)=\sum_{i<j}w_{ij}(h_i-h_j)e_i\wedge e_j,
 \label{eq:cobracket_difference}
\end{equation}
so the graph difference appears as the coefficient of the oriented edge basis.

\begin{lemma}\label{lem:dualbracket}
Let $\{e^a\}$ be the dual basis.  The bracket on $\mathfrak g^*$ induced by $\delta$ is
\begin{equation}
 [e^a,e^b]^*=w_{ab}(e^a-e^b).
 \label{eq:dual_bracket}
\end{equation}
Consequently the dual bracket has the same structure constants as the original bracket and satisfies Jacobi's identity under the same weight conditions.
\end{lemma}
\textbf{Proof}: By duality,
$\langle[e^a,e^b]^*,e_c\rangle=\langle e^a\wedge e^b,\delta(e_c)\rangle$.
Using eq.~(\ref{eq:cobracket_basis}) gives
$w_{cb}\delta^a_c-w_{ca}\delta^b_c$, which is precisely the pairing of $w_{ab}(e^a-e^b)$ with $e_c$. \hfill Q.E.D.

\begin{theorem}[Exact canonical Lie-bialgebra compatibility]\label{thm:cocycle}
Let $R$ be a commutative unital ring without nonzero $2$-torsion, let $w_{ij}=w_{ji}$, and fix the edge-difference cobracket~(\ref{eq:cobracket_basis}) with these same weights.  Then the bracket and this cobracket define a Lie bialgebra if and only if the incident-annihilation condition~(\ref{eq:incident_annihilation}) holds.  In the sufficient direction, the cobracket satisfies the co-Jacobi identity and the $1$-cocycle condition
\begin{equation}
 \delta([x,y])=
 \operatorname{ad}^{(2)}_x\delta(y)
 -\operatorname{ad}^{(2)}_y\delta(x),
 \label{eq:cocycle_identity}
\end{equation}
where
$\operatorname{ad}^{(2)}_x(u\wedge v)=[x,u]\wedge v+u\wedge[x,v]$.
Thus $(\mathfrak g,[\ ,\ ],\delta)$ is a Lie bialgebra.
\end{theorem}
\textbf{Proof}: The co-Jacobi identity is equivalent to Jacobi's identity for the dual bracket and follows from Lemma~\ref{lem:dualbracket}.  It remains to prove eq.~(\ref{eq:cocycle_identity}).  By bilinearity it is enough to take $x=e_a$, $y=e_b$.  Put $u=w_{ab}$ and, for $c\neq a,b$, put $p_c=w_{ac}$ and $q_c=w_{bc}$.  The incident-annihilation relations give
\begin{equation}
 up_c=uq_c=p_cq_c=0.
 \label{eq:cocycle_annihilations}
\end{equation}
Since $[e_a,e_b]=u(e_a-e_b)$,
\begin{equation}
\begin{split}
 \delta([e_a,e_b])
={}&u\bigl(\delta(e_a)-\delta(e_b)\bigr)\\
={}&2u^2e_a\wedge e_b
 +\sum_{c\neq a,b}
 \left(up_c e_a\wedge e_c-uq_c e_b\wedge e_c\right)\\
={}&2u^2e_a\wedge e_b.
\end{split}
\label{eq:cocycle_lhs}
\end{equation}
For the first term on the right-hand side, the $ab$ contribution is
\begin{equation}
 \operatorname{ad}^{(2)}_{e_a}(-u e_a\wedge e_b)=u^2e_a\wedge e_b.
\end{equation}
Every remaining term has the form
\begin{equation}
 q_c\left(u(e_a-e_b)\wedge e_c+p_c e_b\wedge(e_a-e_c)\right)
\end{equation}
and vanishes by eq.~(\ref{eq:cocycle_annihilations}).  Hence
\begin{equation}
 \operatorname{ad}^{(2)}_{e_a}\delta(e_b)=u^2e_a\wedge e_b.
 \label{eq:cocycle_rhs1}
\end{equation}
Similarly,
\begin{equation}
 \operatorname{ad}^{(2)}_{e_b}\delta(e_a)=-u^2e_a\wedge e_b,
 \label{eq:cocycle_rhs2}
\end{equation}
while all terms involving a third vertex again vanish by eq.~(\ref{eq:cocycle_annihilations}).  Subtracting eq.~(\ref{eq:cocycle_rhs2}) from eq.~(\ref{eq:cocycle_rhs1}) reproduces eq.~(\ref{eq:cocycle_lhs}). \hfill Q.E.D.

\textbf{Necessity.} Assume the bracket and the prescribed cobracket satisfy the $1$-cocycle identity.  Fix three distinct vertices $a,b,c$ and write $u=w_{ab}$, $v=w_{ac}$, and $z=w_{bc}$ (zero values allowed).  Compare the coefficients of the oriented wedges supported on $\{a,b,c\}$ in the cocycle defects for the three basis pairs $(e_a,e_b)$, $(e_a,e_c)$, and $(e_b,e_c)$.  For example, define the cocycle defect $\mathcal E_{ab}=\delta([e_a,e_b])-\operatorname{ad}^{(2)}_{e_a}\delta(e_b)+\operatorname{ad}^{(2)}_{e_b}\delta(e_a)$. Direct expansion using~(\ref{eq:e_com_rel}) and~(\ref{eq:cobracket_basis}) gives $[e_a\wedge e_c]\mathcal E_{ab}=-z(u+v)$ and $[e_b\wedge e_c]\mathcal E_{ab}=v(u+z)$, where brackets mean ``coefficient of''. Requiring these coefficients to vanish, and using the cyclic basis-pair defects, yields
\begin{equation}
 u(v+z)=0,\qquad v(u+z)=0,\qquad z(u+v)=0.
 \label{eq:cocycle_necessity_triple}
\end{equation}
Terms from edges meeting other vertices carry those other basis labels and cannot contribute to the displayed wedge coefficients.  Adding the first two identities and subtracting the third gives $2uv=0$; cyclically, $2uz=2vz=0$.  Since $R$ has no nonzero $2$-torsion, $uv=uz=vz=0$.  Every pair of distinct incident edges lies in some such triple, hence~(\ref{eq:incident_annihilation}) follows. \hfill Q.E.D.

The absence of nonzero $2$-torsion is essential to this converse; it does not require $2$ to be invertible.  For example, over $\mathbb F_2$ a triangle with all three weights equal to $1$ satisfies Jacobi's identity and the canonical cocycle identity despite products of incident edge weights being nonzero.  

The co-Jacobi identity follows because the coefficients defining the cobracket also define a Lie bracket on the dual space. The direct calculation above establishes the independent $1$-cocycle condition.  Up to orientation convention, eq.~(\ref{eq:cobracket_difference}) is the weighted graph difference operator.  The bracket and cobracket therefore form a closed algebraic structure in which the graph Laplacian is generated internally by $\operatorname{ad}_1=-\Delta_w$.

\begin{remark}[Scope of the bialgebra construction]
For $R=\mathbb C^q$, restriction to each idempotent channel gives an ordinary finite-dimensional complex Lie bialgebra and so the standard Drinfeld double can be constructed separately in each coefficient component.  The double is not required for the transport results below and will not be developed further.  The structures used in the applications are the inner derivative $\operatorname{ad}_1$, the edge differential $\delta$, and the second-order operator $\delta^\dagger\delta$.
\end{remark}

The cobracket records weighted boundary edges separately in each coefficient component.  Let $S\subset V(G)$, and let $\chi_S$ denote the characteristic function of $S$.  Writing $w_{ij}=\sum_\alpha w_{ij}^{(\alpha)}\varepsilon_\alpha$, one obtains $\delta(\chi_S) = \sum_\alpha \varepsilon_\alpha\delta_\alpha(\chi_S)$, where $\delta_\alpha(\chi_S) = \sum_{i<j} w_{ij}^{(\alpha)} \bigl(\chi_S(i)-\chi_S(j)\bigr)e_i\wedge e_j$.

Thus $\delta_\alpha(\chi_S)$ records precisely the weighted boundary edges of $S$ that belong to the $\alpha$-th idempotent channel. Equivalently, only edges crossing the cut $(S,V\setminus S)$ contribute to $\delta(\chi_S)$, and the idempotent decomposition separates this boundary into channel components, so the cobracket gives the coalgebraic form of the same separation into matching-supported components that appears in the dynamics.

More precisely, the cobracket does not merely identify the boundary of a subset. It records the oriented, channel-resolved edge contributions across that boundary.

\begin{corollary}[Complete matching-layer characterization over $\mathbb C^q$]
\label{cor:complete_matching_layers}
Let $w_e=\sum_{\alpha=1}^q\gamma_{e,\alpha}\varepsilon_\alpha$ be arbitrary symmetric weights over $\mathbb C^q$. The prescribed canonical edge-difference cobracket is compatible with the graph bracket if and only if, for each $\alpha$, the set of edges with $\gamma_{e,\alpha}\ne0$ is a matching. Several different idempotent components may be nonzero on the same edge.
\end{corollary}
\textbf{Proof.} Theorem~\ref{thm:cocycle} is equivalent to $\gamma_{e,\alpha}\gamma_{f,\alpha}=0$ for any distinct incident $e,f$ and every $\alpha$. Because $\mathbb C$ is a domain, no two such coefficients can both be nonzero. The converse is immediate. \hfill Q.E.D.

\begin{remark}[Nilpotent edge weights]
\label{ex:dual_number}
Consider the ring $R=\mathbb C[x]/(x^2)$ and assign the weight $w_e=x$ to every edge of an arbitrary finite simple graph. Since $x^2=0$, Jacobi's identity and the compatibility condition for the specified cobracket hold, even when several edges meet at a vertex. Thus the matching restriction does not apply to this ring.

The resulting complex Lie algebra is two-step nilpotent (or abelian for an edgeless graph), since $[e_i,e_j]=x(e_i-e_j)$ on edges and $[xR^V,R^V]=0$. It is a quotient of the graph Lie algebra considered by Farinati and Jancsa~\cite{farinati}: their vertex generators correspond to $e_i$, and their central edge generators to $x(e_i-e_j)$. This correspondence concerns the Lie brackets, not the cobrackets.

This example also explains why we use nonnilpotent edge weights in the physical applications: for the weights considered here, the first-order edge operator may be nonzero, but its second-order counterpart vanishes because $x^2=0$.
\end{remark}

The following factorization identifies the precise role of the bialgebra in the quantum and stochastic applications.

\subsection{Bialgebraic differential factorization}\label{sec:bialgebra_factorization}
For the remainder of the paper let the coefficient algebra be either $\mathcal A=\mathbb C^q$ with its componentwise involution or its ordered self-adjoint part $R=\mathbb R^q$, and assume symmetric self-adjoint weights.  Orient each edge by $i<j$ and equip the edge module with the orthonormal basis $e_i\wedge e_j$.  Define the unweighted incidence boundary
\begin{equation}
 \partial(e_i\wedge e_j)=e_i-e_j,\qquad i<j,
 \label{eq:edge_boundary}
\end{equation}
and let $\delta^\dagger$ be the adjoint of $\delta$ for the standard vertex and edge inner products.

\begin{theorem}[Bracket--cobracket factorization]\label{thm:bialgebra_factorization}
For every vertex function $f$,
\begin{equation}
 \Delta_w f=\partial\delta f=-[1,f].
 \label{eq:first_order_factorization}
\end{equation}
Moreover,
\begin{equation}
 \delta^\dagger\delta f=\Delta_{w^2}f
 :=\sum_j w_{ij}^2(f_i-f_j).
 \label{eq:cobracket_laplacian}
\end{equation}
Under the incident-annihilation condition~(\ref{eq:incident_annihilation}),
\begin{equation}
 \Delta_w^2=2\Delta_{w^2}=2\delta^\dagger\delta.
 \label{eq:squared_laplacian_factorization}
\end{equation}
\end{theorem}
\textbf{Proof}: The identities must first be checked on the full weighted graph, before separating its idempotent components. Let $e_i$ denote the characteristic function of vertex $i$, put $w_{ii}=0$, and write $\mathfrak d_i=\sum_{k\ne i}w_{ik}$. Directly from the definition,
\[
 (\Delta_w e_i)(i)=\mathfrak d_i=\sum_{k\sim i}w_{ik},\qquad
 (\Delta_w e_i)(j)=-w_{ji}\quad(j\ne i).
\]
In particular, the diagonal entry is the sum of all weights incident at $i$, not the weight of one arbitrarily selected edge. The full Laplacian matrix satisfies $(\Delta_w)_{ii}=\mathfrak d_i$ and $(\Delta_w)_{ij}=-w_{ij}$ for $i\ne j$. Matrix multiplication without any matching assumption therefore gives
\[
 (\Delta_w^2)_{ii}=\mathfrak d_i^2+\sum_{k\ne i}w_{ik}^{2},\qquad
 (\Delta_w^2)_{ij}=-w_{ij}(\mathfrak d_i+\mathfrak d_j)
                  +\sum_{k\notin\{i,j\}}w_{ik}w_{kj}\quad(i\ne j).
\]
Now impose incident annihilation. Commutativity implies $\mathfrak d_i^2=\sum_kw_{ik}^2$, $w_{ij}\mathfrak d_i=w_{ij}\mathfrak d_j=w_{ij}^2$, and every summand $w_{ik}w_{kj}$ with $i\ne j$ and $k\notin\{i,j\}$ vanishes. Thus
\[
 (\Delta_w^2)_{ii}=2\sum_kw_{ik}^2,\qquad
 (\Delta_w^2)_{ij}=-2w_{ij}^2\quad(i\ne j),
\]
which are exactly the entries of $2\Delta_{w^2}$. Independently, applying the unweighted boundary $\partial$ to the oriented edge differential $\delta f$ gives $\Delta_w f$, and its adjoint multiplies each edge difference by the second (self-adjoint) weight, so that $\delta^\dagger\delta=\Delta_{w^2}$. Finally $[1,f]=-\Delta_wf$ by the bracket definition. Together these prove all three identities. \hfill Q.E.D.

\noindent\textbf{Spectral interpretation.} For $\mathcal A=\mathbb C^q$ or $R=\mathbb R^q$, resolve the coefficient weights as $w_e=\sum_{\alpha}w_e^{(\alpha)}\varepsilon_\alpha$. In each component of the coefficient ring, the weight-determined support $\mathcal M_\alpha=\{e:w_e^{(\alpha)}\ne0\}$ is a matching. The channel Laplacian is therefore a direct sum of zero blocks for unmatched vertices and two-vertex blocks
\[
 \Delta_e^{(\alpha)}=w_e^{(\alpha)}
 \begin{pmatrix}1&-1\\-1&1\end{pmatrix},\qquad
 \operatorname{spec}(\Delta_e^{(\alpha)})=\{0,2w_e^{(\alpha)}\}.
\]
The endpoint vectors $e_i,e_j$ are not eigenvectors when $w_e^{(\alpha)}\ne0$; the symmetric and antisymmetric combinations $(e_i\pm e_j)/\sqrt2$ are. Squaring each block replaces its nonzero eigenvalue $2w_e^{(\alpha)}$ by $4(w_e^{(\alpha)})^2$, precisely twice the corresponding eigenvalue of $\Delta_{(w^{(\alpha)})^2}$. This explains the factor $2$ in eq.~(\ref{eq:squared_laplacian_factorization}) and the nontrivial motion confined to each two-vertex cage. The spectral interpretation uses the self-adjoint representation of the coefficient algebra. The preceding calculation establishes the polynomial identity directly on the full graph, without choosing a matching decomposition or an eigenbasis. Here $\Delta_w^2$ denotes operator composition, whereas $\Delta_{w^2}$ denotes the Laplacian with squared edge weights.\begin{corollary}[Common transport operator]\label{cor:common_transport_operator}
With the conventions used below,
\begin{align}
 D&=i\Delta_w=i\partial\delta=-i\operatorname{ad}_1,
 &\hat p&=-i\hbar D=\hbar\Delta_w=-\hbar\operatorname{ad}_1,
 \label{eq:quantum_bialgebra_bridge}\\
 \hat T&=\frac{\hat p^2}{2m}
 =\frac{\hbar^2}{m}\delta^\dagger\delta,
 \label{eq:quantum_kinetic_cobracket}\\
 L&=-\frac12\Delta_w=\frac12\operatorname{ad}_1,
 &L^2&=\frac12\delta^\dagger\delta.
 \label{eq:FP_bialgebra_bridge}
\end{align}
Consequently the graph Fokker--Planck equation~(\ref{eq:FPE_ordered_ring}) has the bialgebraic divergence form
\begin{equation}
 \partial_tP
 =\frac12\partial\delta(AP)
 +\frac12\delta^\dagger\delta(BP)
 =-\frac12[1,AP]+\frac12\delta^\dagger\delta(BP).
 \label{eq:FPE_bialgebra_form}
\end{equation}
\end{corollary}

The cocycle identity is the compatibility law between the vertex bracket and the edge differential $\delta$.  Put $Q=\delta^\dagger\delta$, a nonnegative operator in every self-adjoint idempotent component.  The ring-valued identities must be resolved channelwise before interpreting operators on an ordinary Hilbert space.  In particular, a scalar averaging functional $\varphi$ on $\mathbb C^q$ is generally not multiplicative: $\varphi(\Delta_w^2)$ need not equal $[\varphi(\Delta_w)]^2$.  Averaging channel weights first can join adjacent edges and destroy the matching-block spectrum; a physical mixture of channels must instead be interpreted using the corresponding channel evolutions and state/probability weights.  Equations~(\ref{eq:quantum_kinetic_cobracket}) and~(\ref{eq:FP_bialgebra_bridge}) identify the common algebraic factor $Q$, but not a common physical time-evolution generator.  Indeed, with $A=0$ and $B=-2\operatorname{diag}(\nu_v)$ for componentwise nonnegative $\nu_v$, the stochastic generator in eq.~(\ref{eq:FPE_ordered_ring}) becomes
\begin{equation}
 M=-Q\operatorname{diag}(\nu_v),\qquad M_{ij}=w_{ij}^2\nu_j\quad(i\ne j),
 \label{eq:physical_generator_distinction}
\end{equation}
whereas the quantum kinetic operator is $\hat T=(\hbar^2/m)Q$.  In the homogeneous case $\nu_v=\nu$ these obey $M=-(m\nu/\hbar^2)\hat T$.  The sign, the coefficients $\nu_v$, and the conditions for stochastic positivity are physical inputs, not consequences of the Lie-bialgebra axioms.  General $A,B$ must separately satisfy the off-diagonal Markov inequalities~(\ref{eq:admissible_B_condition}).

\section{Applications}

We next apply Corollary~\ref{cor:common_transport_operator} to coherent and stochastic dynamics.  The common nonnegative second-order edge operator is $Q=\delta^\dagger\delta$.  The graph operator is initially defined using the full weighted degree $\mathfrak d_i=\sum_jw_{ij}$, so its vertex-basis matrix is not generally diagonal.  The orthogonal two-vertex spectral blocks arise only after resolving the weight-determined matching in each idempotent channel; no matching is chosen independently of the weights.  Quantum dynamics uses a positive multiple of $Q$ as kinetic energy; pure-diffusion stochastic dynamics uses $-Q\operatorname{diag}(\nu_v)$ as its Markov generator.  The products of weights on distinct incident edges vanish, whereas the squares of nonzero edge weights do not.  Hermiticity and Markov positivity are separately imposed; fixed-channel caging additionally requires the physical hopping or jump rates to be supported on the corresponding matching.

\subsection{Schr\"odinger equation over an ordered complex zero-divisor algebra}
\label{subsec:Schrodinger_ordered_complex}

The bracket--cobracket factorization in Theorem~\ref{thm:bialgebra_factorization} supplies the discrete derivative and its positive second-order form.  For the Schr\"odinger equation these operators must also be compatible with complex conjugation, adjoints, and the positivity of probabilities. We therefore use the finite commutative complex $*$-algebra
\begin{equation}
        {\mathcal A}={\mathbb C}^q,
        \qquad q\geq 1,
        \label{eq:complex_ordered_algebra}
\end{equation}
with componentwise addition and multiplication. The involution is componentwise
complex conjugation,
\begin{equation}
        z^*=\sum_{\alpha=1}^q \overline{z^{(\alpha)}}\varepsilon_\alpha,
        \qquad z=\sum_{\alpha=1}^q z^{(\alpha)}\varepsilon_\alpha.
\end{equation}
Here $\varepsilon_1,\ldots,\varepsilon_q$ are the standard idempotents,
\begin{equation}
        \varepsilon_\alpha\varepsilon_\beta=0\quad(\alpha\neq\beta),
        \qquad
        \varepsilon_\alpha^2=\varepsilon_\alpha,
        \qquad
        \sum_{\alpha=1}^q\varepsilon_\alpha=1_{\mathcal A}.
\label{eq:idempotents_complex}
\end{equation}
Only the self-adjoint part of a complex algebra is ordered. We write
\begin{equation}
        {\mathcal A}_{\rm h}=\{z\in{\mathcal A}:z=z^*\}\simeq {\mathbb R}^q
\end{equation}
and define
\begin{equation}
        x\geq 0
        \quad\Longleftrightarrow\quad
        x=x^*,\quad x^{(\alpha)}\geq0\quad\textrm{for every }\alpha.
\label{eq:complex_positive_cone}
\end{equation}
Equivalently,
\begin{equation}
        {\mathcal A}_+
        =\{z^*z:z\in{\mathcal A}\}
        =\left\{\sum_\alpha x^{(\alpha)}\varepsilon_\alpha:
        x^{(\alpha)}\geq0\right\}.
\end{equation}
Thus the modulus is defined by
\begin{equation}
        |z|^2=z^*z\in {\mathcal A}_+ .
\end{equation}
This is the finite-dimensional commutative $C^*$-algebra
$C(\{1,\ldots,q\},{\mathbb C})$.  It is the complex counterpart of the
ordered real function algebras used in lattice-ordered ring theory and in the
Fokker--Planck subsection. Positive elements of a $C^*$-algebra are
standardly defined through the $*$-structure; see, for example,
\cite{arveson,kadison_ringrose}. The real ordered part is the algebra
${\mathbb R}^q$ discussed in \cite{birkhoff_pierce,gj,zaanen,ab_positive}. A nonzero element $z\in {\mathcal A}$ is a zero divisor if there exists a
nonzero element $y\in {\mathcal A}$ such that $zy=0$. For $q>1$, the
idempotents $\varepsilon_\alpha$ are non-nilpotent zero divisors:
\begin{equation}
        \varepsilon_\alpha(1_{\mathcal A}-\varepsilon_\alpha)=0,
        \qquad
        \varepsilon_\alpha^2=\varepsilon_\alpha\neq0.
\end{equation}
Hence distinct components may annihilate each other, while the square of one
component survives. This is the zero-divisor feature that is needed in the
applications below.

\subsubsection{Edge colours and the skew-adjoint graph derivative}

Let $G$ be an undirected graph with vertices $1,\ldots,N$. To each unoriented edge $j - k$ assign a colour
$c(j,k)=c(k,j)\in\{1, \ldots,q\}$ and a real symmetric amplitude
$\gamma_{jk}=\gamma_{kj}>0$ on edges, and set $\gamma_{jk}=0$ for non-edges. The positive ordered edge weight is
\begin{equation}
        w_{jk}=\gamma_{jk}\varepsilon_{c(j,k)}\in {\mathcal A}_+,
        \qquad w_{jk}=w_{kj}.
\label{eq:ordered_edge_weight}
\end{equation}
We impose the local colouring condition
\begin{equation}
        c(j,k)\neq c(j,l)
        \quad\textrm{whenever }k\neq l,
        \quad j\sim k,
        \quad j\sim l.
\label{eq:local_colour_condition}
\end{equation}
Consequently,
\begin{equation}
        w_{jk}w_{jl}=0,
        \qquad k\neq l,
        \label{eq:zero_product_incident}
\end{equation}
whereas
\begin{equation}
        w_{jk}^2=\gamma_{jk}^2\varepsilon_{c(j,k)}\in {\mathcal A}_+.
\label{eq:square_survives}
\end{equation}
The product of two distinct incident edge weights therefore vanishes, but an
individual edge weight is not square-zero unless $\gamma_{jk}=0$. For quantum mechanics the first-order derivative should be skew-adjoint.  We therefore use the purely imaginary graph derivative
\begin{equation}
        (D\Psi)_j
        =\sum_k \omega_{jk}(\Psi_j-\Psi_k),
        \qquad
        \omega_{jk}=i\,w_{jk}.
\label{eq:skew_graph_derivative}
\end{equation}

Since $w_{jk}=w_{kj}=w_{jk}^*$, one has
\begin{equation}
        \omega_{jk}=\omega_{kj},
        \qquad
        \omega_{jk}^*=-\omega_{kj}.
\end{equation}
Let
\begin{equation}
        \langle \Phi,\Psi\rangle
        =\sum_j \Phi_j^*\Psi_j
        \in {\mathcal A}
        \label{eq:A_inner_product}
\end{equation}
be the ${\mathcal A}$-valued inner product. A direct summation using
$\omega_{ij}=\omega_{ji}$ and $\omega_{ij}^*=-\omega_{ij}$ gives
\begin{equation}
        \langle \Phi,D\Psi\rangle
        =-\langle D\Phi,\Psi\rangle .
\label{eq:D_skew_adjoint}
\end{equation}
Thus $D$ is skew-adjoint.

\subsubsection{Spectral decomposition of the graph operator}

The momentum operator associated with the graph derivative is
\begin{equation}
        \hat p=-i\,\hbar D .
\label{eq:momentum_ordered_complex}
\end{equation}
By (\ref{eq:D_skew_adjoint}),
\begin{equation}
        \hat p^*=\hat p,
\end{equation}
so $\hat p$ is Hermitian. The kinetic energy is
\begin{equation}
        \hat T=\frac{\hat p^2}{2m}
        =-\frac{\hbar^2}{2m}D^2
        =\frac{\hbar^2}{m}\delta^\dagger\delta,
\label{eq:kinetic_ordered_complex}
\end{equation}
where the final identity is Corollary~\ref{cor:common_transport_operator}.  Thus the cobracket is the channel-resolved quantum gradient and its adjoint supplies the corresponding divergence.  In particular,
\begin{equation}
 \langle\Psi,\hat T\Psi\rangle
 =\frac{\hbar^2}{m}\langle\delta\Psi,\delta\Psi\rangle_E
 =\frac{\hbar^2}{m}\sum_{i<j}w_{ij}^2|\Psi_i-\Psi_j|^2
 \in\mathcal A_+.
 \label{eq:kinetic_edge_energy}
\end{equation}
It is also positive in the ordered $*$-algebraic sense directly from the momentum representation, because
\begin{equation}
        \langle \Psi,\hat T\Psi\rangle
        =\frac{1}{2m}\langle \hat p\Psi,\hat p\Psi\rangle
        =\frac{1}{2m}\sum_j |(\hat p\Psi)_j|^2
        \in {\mathcal A}_+ .
\label{eq:T_positive_modulus}
\end{equation}
The operator $\hat p=\hbar\Delta_w$ is self-adjoint and has a nonnegative spectrum for positive edge weights. Since the edges with nonzero weights form a matching in each channel, the operator decomposes into independent one- and two-vertex blocks. On an edge $\{i,j\}$, the corresponding eigenstates are the symmetric and antisymmetric combinations of the endpoint states. Thus a particle confined to a fixed channel cannot propagate beyond a single edge. For $\mathcal A=\mathbb C^q$, the spectral problem separates into $q$ ordinary finite-dimensional complex problems. Writing
\begin{equation}
        D=\sum_{\alpha=1}^q \varepsilon_\alpha D^{(\alpha)},
        \qquad
        \hat p=\sum_{\alpha=1}^q \varepsilon_\alpha \hat p^{(\alpha)},
\end{equation}
each $\hat p^{(\alpha)}$ is an ordinary Hermitian matrix. Hence there is an orthonormal eigenbasis
\begin{equation}
        \hat p^{(\alpha)}\phi_r^{(\alpha)}
        =p_r^{(\alpha)}\phi_r^{(\alpha)},
        \qquad p_r^{(\alpha)}\in {\mathbb R}.
\label{eq:graph_fourier_modes}
\end{equation}
The expansion
\begin{equation}
        \Psi
        =\sum_{\alpha=1}^q\varepsilon_\alpha
        \sum_r a_r^{(\alpha)}\phi_r^{(\alpha)}
        \label{eq:graph_fourier_expansion}
\end{equation}
is the graph Fourier expansion associated with the derivative $D$.

\subsubsection{Schr\"odinger equation and probability conservation}

Let the potential be a self-adjoint multiplication operator,
\begin{equation}
        (\hat U\Psi)_k=U_k\Psi_k,
        \qquad U_k=U_k^*\in {\mathcal A}_{\rm h}.
\end{equation}
The Hamiltonian is
\begin{equation}
        \hat H=\hat T+\hat U.
\end{equation}
Both $\hat T$ and $\hat U$ are Hermitian, hence $\hat H$ is Hermitian.
The Schr\"odinger equation is
\begin{equation}
        i\,\hbar\frac{d\Psi}{dt}=\hat H\Psi .
\label{eq:Sch_ordered_complex}
\end{equation}
The probability density at a vertex is
\begin{equation}
        P_k=|\Psi_k|^2=\Psi_k^*\Psi_k\in {\mathcal A}_+ .
\label{eq:node_probability_modulus}
\end{equation}
The total probability is
\begin{equation}
        \|\Psi\|^2=\langle\Psi,\Psi\rangle=
        \sum_k |\Psi_k|^2\in {\mathcal A}_+.
\end{equation}
A normalised state satisfies $\|\Psi\|^2=1_{\mathcal A}$.  If a scalar real probability is desired, one may apply any positive state 
$\varphi:{\mathcal A}\rightarrow {\mathbb C}$, for example
$\varphi(\sum_\alpha z^{(\alpha)}\varepsilon_\alpha)=
\sum_\alpha \lambda_\alpha z^{(\alpha)}$, with
$\lambda_\alpha\geq0$ and $\sum_\alpha\lambda_\alpha=1$. Then $\varphi(P_k)\geq0$ and $\sum_k\varphi(P_k)=1$.

The evolution generated by (\ref{eq:Sch_ordered_complex}) conserves the total
probability. Indeed,
\begin{eqnarray}
        \frac{d}{dt}\langle\Psi,\Psi\rangle
        &=&\langle \dot\Psi,\Psi\rangle+
        \langle\Psi,\dot\Psi\rangle                                                    \\
        &=&\frac{i}{\hbar}\langle \hat H\Psi,\Psi\rangle
        -\frac{i}{\hbar}\langle\Psi,\hat H\Psi\rangle=0,
\end{eqnarray}
because $\hat H=\hat H^*$. This is the global conservation law.

The local conservation law follows by computing the time derivative of
$P_k$. The potential terms cancel pointwise, because $U_k=U_k^*$.  With
$\hat T=-\hbar^2D^2/(2m)$, one obtains
\begin{equation}
        \frac{dP_k}{dt}
        =-\frac{i\hbar}{2m}
        \left((D^2\Psi)_k^*\Psi_k-
        \Psi_k^*(D^2\Psi)_k\right).
\label{eq:local_density_derivative}
\end{equation}
This expression can be written as a divergence.  Define the oriented current
from $k$ to $j$ by
\begin{eqnarray}
        J_{kj}&=&-\frac{i\hbar}{2m}\,\omega_{kj}
        \Big(\Psi_j^*(D\Psi)_k-
        \Psi_k^*(D\Psi)_j \nonumber \\
        &&\hspace{2.2cm}
        +\Psi_k(D\Psi^*)_j-
        \Psi_j(D\Psi^*)_k\Big).
\label{eq:Sch_current_ordered_complex}
\end{eqnarray}
Here $D\Psi^*$ means that the same difference operator $D$ is applied to
$\Psi^*$. The current is antisymmetric,
\begin{equation}
        J_{jk}=-J_{kj},
\end{equation}
and a direct substitution of
$(D^2\Psi)_k=\sum_j\omega_{kj}((D\Psi)_k-(D\Psi)_j)
$ gives
\begin{equation}
        \frac{dP_k}{dt}+\sum_j J_{kj}=0.
\label{eq:Sch_continuity_ordered_complex}
\end{equation}
Thus probability is locally conserved. The statement is stronger than global conservation: probability can change at a node only through antisymmetric edge currents. Using the zero-product rule (\ref{eq:zero_product_incident}), the current along
one edge simplifies. Since products involving two different incident edge weights vanish, only the same-edge square remains, and for $k\sim j$
\begin{equation}
        J_{kj}
        =-\frac{i\hbar}{m}\,\omega_{kj}^2
        \left(\Psi_j^*\Psi_k-\Psi_k^*\Psi_j\right).
\label{eq:Sch_current_simplified_omega}
\end{equation}
Because $\omega_{kj}=i\,w_{kj}$, this is equivalently
\begin{equation}
        J_{kj}
        =\frac{i\hbar}{m}\,w_{kj}^2
        \left(\Psi_j^*\Psi_k-\Psi_k^*\Psi_j\right).
\label{eq:Sch_current_simplified_w}
\end{equation}
If, in one component, $\Psi_k=\sqrt{P_k}\exp(i\theta_k)$ and $\Psi_j=\sqrt{P_j}\exp(i\theta_j)$, then
\begin{equation} 
        J_{kj}
        =\frac{2\hbar}{m}\,w_{kj}^2
        \sqrt{P_kP_j}\,\sin(\theta_j-\theta_k)
        \label{eq:Sch_current_phase}
\end{equation}
with the convention of (\ref{eq:Sch_continuity_ordered_complex}). Thus the current is controlled by the edge weight $w_{kj}^2$ and by the phase difference across the edge.  It is generally nonzero. It vanishes only for zero phase difference modulo $\pi$, for a zero amplitude at one endpoint, or under the stronger square-zero condition $w_{kj}^2=0$. The zero-divisor condition (\ref{eq:zero_product_incident}) kills mixed two-edge terms; it does not kill the same-edge current. \subsubsection{Explicit free-particle solution}

For a free particle $\hat U=0$, and the equation is
\begin{equation}
        i\,\hbar\frac{d\Psi}{dt} =\frac{\hat p^2}{2m}\Psi .
\label{eq:free_Sch_ordered_complex}
\end{equation}
Using the graph Fourier expansion (\ref{eq:graph_fourier_expansion}) and the eigenvalue equation (\ref{eq:graph_fourier_modes}), the solution is 
\begin{equation}
        \Psi(t) = \sum_{\alpha=1}^q\varepsilon_\alpha \sum_r a_r^{(\alpha)} \exp\left(-\frac{i}{\hbar}E_r^{(\alpha)}t\right) \phi_r^{(\alpha)}, \label{eq:free_solution_spectral}
\end{equation}
where
\begin{equation}
        E_r^{(\alpha)}=\frac{(p_r^{(\alpha)})^2}{2m}\geq0.
\label{eq:free_energy_spectral}
\end{equation}
The normalisation condition is
\begin{equation}
        \sum_{\alpha=1}^q\varepsilon_\alpha \sum_r |a_r^{(\alpha)}|^2=1_{\mathcal A}.
\end{equation}
Equation~(\ref{eq:free_solution_spectral}) is the finite-dimensional eigenmode expansion of the graph Hamiltonian. Translation symmetry may permit alternative extended eigenbases inside degenerate eigenspaces, but cannot produce transport between distinct matching components within a fixed channel. The strict zero-divisor colouring gives an even more explicit form. In a fixed colour component $\alpha$, the edges of colour $\alpha$ form a matching. Hence each connected component is either an isolated vertex or a single edge. For an edge $a - b$ of colour $\alpha$ with weight $w_{ab}=\gamma_{ab}\varepsilon_\alpha$, the two normalised eigenvectors are 
\begin{equation}
        \phi_{ab,+}^{(\alpha)}=\frac{e_a+e_b}{\sqrt{2}}, \qquad \phi_{ab,-}^{(\alpha)}=\frac{e_a-e_b}{\sqrt{2}}.
\end{equation}
They satisfy 
\begin{equation}
        \hat p^{(\alpha)}\phi_{ab,+}^{(\alpha)}=0, \qquad \hat p^{(\alpha)}\phi_{ab,-}^{(\alpha)}=2\hbar\gamma_{ab} \phi_{ab,-}^{(\alpha)}. 
\end{equation}
Therefore
\begin{equation}
        E_{ab,+}^{(\alpha)}=0, \qquad  E_{ab,-}^{(\alpha)}=\frac{(2\hbar\gamma_{ab})^2}{2m} = \frac{2\hbar^2\gamma_{ab}^2}{m}.
\end{equation}
The free solution on this edge is
\begin{eqnarray}
        \Psi_a^{(\alpha)}(t)&=&\frac{1}{\sqrt2} \left(c_+ + c_-\exp(-i\Omega_{ab}t)\right), \\
        \Psi_b^{(\alpha)}(t)&=&\frac{1}{\sqrt2} \left(c_+ - c_-\exp(-i \Omega_{ab}t)\right),
        \label{eq:free_edge_solution}
\end{eqnarray}
where
\begin{equation}
        \Omega_{ab}=\frac{E_{ab,-}^{(\alpha)}}{\hbar}
        =\frac{2\hbar\gamma_{ab}^2}{m}.
\end{equation}
A single energy eigenmode has time-independent density.  A superposition of the symmetric and antisymmetric modes has a nontrivial relative phase and therefore may have a nonzero current along the edge, exactly as predicted by
(\ref{eq:Sch_current_phase}). This behaviour is physically consistent with the imposed algebraic constraint. Hermiticity gives unitary evolution, the kinetic energy is positive in the modulus sense, and local probability is conserved. The zero-divisor condition prevents fixed-channel transport between distinct matching components. A basis of spatially extended linear combinations can exist within a degenerate eigenvalue, but no such choice of basis changes this dynamical confinement. Thus the eigenvalues of $\hat p$ are graph-spectral labels, not ordinary spatial momenta.
Some remarks have to be made on relation to existing literature on quantum graphs. The present construction is not a metric quantum graph. In a quantum graph, the edges are one-dimensional continua and the Schrödinger operator is a differential operator acting along these metric edges. Here the wavefunction is defined on the vertex set and the derivative is an algebraic graph operator with coefficients in an ordered zero-divisor algebra. No embedding of the graph in an ambient Euclidean space and no wire-like physical medium are assumed. The coefficient algebra supplies an inner derivation of the graph-function Lie bracket, not a derivation of the pointwise product; under incident annihilation, its square is proportional to the positive edge Laplacian used for kinetic energy.

\subsection{Quantum caging and entangling routing}\label{sec:caging_router}

Let
\begin{equation}
 {\mathcal A}=\mathbb C^q=\bigoplus_{\alpha=1}^q\mathbb C\varepsilon_\alpha,
 \qquad
 \varepsilon_\alpha\varepsilon_\beta=\delta_{\alpha\beta}\varepsilon_\alpha.
 \label{eq:channel_algebra}
\end{equation}
A wavefunction has the channel decomposition
$\Psi_j=\sum_\alpha\psi_j^{(\alpha)}\varepsilon_\alpha$.  If an edge carries the weight $\gamma_{ij}\varepsilon_\alpha$, it acts only on the $\alpha$ component.  A proper edge colouring decomposes every channel graph into a matching.  Hence each channel component evolves on isolated vertices or on independent two-vertex systems.  This produces exact compact localization without phase cancellation: a path that changes colour contains a product $\varepsilon_\alpha\varepsilon_\beta=0$.

\subsubsection{The coloured square}
Consider $C_4$ with vertices $\{1,2,3,4\}$ and alternating weights
\begin{equation}
 w_{12}=w_{34}=\gamma\varepsilon_1,
 \qquad
 w_{23}=w_{41}=\gamma\varepsilon_2,
 \label{eq:caging_square_weights}
\end{equation}
with symmetric reversed-edge weights understood.  The two channel matchings are
$(1-2)\cup(3-4)$ and $(2-3)\cup(4-1)$.  On one matched edge $a-b$, the kinetic Hamiltonian is
\begin{equation}
 H_{ab}=\kappa
 \begin{pmatrix}1&-1\\-1&1\end{pmatrix},
 \qquad
 \kappa=\frac{\hbar^2\gamma^2}{m}.
 \label{eq:caging_edge_hamiltonian}
\end{equation}
Writing $\theta=\kappa t/\hbar$, a component initially at $a$ evolves as
\begin{equation}
 |a\rangle\longmapsto
 a(t)|a\rangle+b(t)|b\rangle,
 \qquad
 a(t)=e^{-i\theta}\cos\theta,
 \quad
 b(t)=i e^{-i\theta}\sin\theta.
 \label{eq:edge_transfer_amplitudes}
\end{equation}
For the algebra-valued initial condition $\Psi(0)=1_{\mathcal A}e_1$, the solution is
\begin{eqnarray}
 \Psi_1(t)&=&a(t)(\varepsilon_1+\varepsilon_2),\nonumber\\
 \Psi_2(t)&=&b(t)\varepsilon_1,\qquad
 \Psi_4(t)=b(t)\varepsilon_2,\nonumber\\
 \Psi_3(t)&=&0.
 \label{eq:caging_square_solution}
\end{eqnarray}
Thus the opposite vertex is dark for all times.  The two shortest scalar paths to vertex $3$ vanish separately because
\begin{equation}
 w_{12}w_{23}=w_{14}w_{43}=0.
 \label{eq:path_annihilation}
\end{equation}
For comparison only, consider the ordinary scalar Laplacian Hamiltonian $H_{\rm sc}=\kappa\Delta_{\rm comb}(C_4)$.  This is a reference Hamiltonian outside the Jacobi-compatible scalar-weight framework: the nonzero scalar-weight support $C_4$ is not a disjoint union of cliques and therefore fails Theorem~\ref{thm:jacobi_exact}.  Its transition probability is
\begin{equation}
 |\psi^{\rm sc}_3(t)|^2=\sin^4(\kappa t/\hbar),
 \label{eq:scalar_square_probability}
\end{equation}
which is generically nonzero.  The localization in eq.~(\ref{eq:caging_square_solution}) is therefore imposed by the coefficient algebra rather than by destructive interference.  It is an algebraic analogue of caging, distinct from the magnetic-flux mechanism of Aharonov--Bohm cages \cite{vidal}.

\subsubsection{Ordinary Hilbert-space realization and entanglement}
To give the channel labels an operational quantum meaning, represent the idempotents on an internal Hilbert space $\mathcal K\cong\mathbb C^q$ by
\begin{equation}
 \varepsilon_\alpha\longmapsto
 P_\alpha=|\alpha\rangle\langle\alpha|.
 \label{eq:idempotent_projectors}
\end{equation}
The algebra-valued Hamiltonian becomes the ordinary Hermitian controlled Hamiltonian
\begin{equation}
 H_{\rm phys}=\sum_{\alpha=1}^q H_\alpha\otimes P_\alpha
 \label{eq:controlled_hamiltonian}
\end{equation}
on $\mathbb C^{|V|}\otimes\mathcal K$.  For the square, prepare the normalized state
\begin{equation}
 |\Phi(0)\rangle=|1\rangle\otimes|+\rangle,
 \qquad
 |+\rangle=\frac{|1_c\rangle+|2_c\rangle}{\sqrt2}.
 \label{eq:router_initial_state}
\end{equation}
Equations (\ref{eq:edge_transfer_amplitudes}) give
\begin{equation}
 |\Phi(t)\rangle=
 \frac{1}{\sqrt2}
 \left[
 (a|1\rangle+b|2\rangle)|1_c\rangle
 +(a|1\rangle+b|4\rangle)|2_c\rangle
 \right].
 \label{eq:router_state}
\end{equation}
At the perfect-transfer time
\begin{equation}
 t_*=\frac{\pi\hbar}{2\kappa},
 \label{eq:transfer_time}
\end{equation}
one obtains, up to an irrelevant global phase,
\begin{equation}
 |\Phi(t_*)\rangle=
 \frac{|2\rangle|1_c\rangle+|4\rangle|2_c\rangle}{\sqrt2}.
 \label{eq:bell_router_state}
\end{equation}
The zero-divisor cage has therefore become a deterministic two-port router and a source of one ebit of path--channel entanglement. One ebit is the entanglement of a maximally entangled pair of two-level quantum systems, corresponding to an entanglement entropy of one bit.

The reduced internal density matrix is
\begin{equation}
 \rho_c(t)=\operatorname{Tr}_{V}|\Phi(t)\rangle\langle\Phi(t)|
 =\frac12
 \begin{pmatrix}
 1&\cos^2\theta\\
 \cos^2\theta&1
 \end{pmatrix}.
 \label{eq:reduced_channel_density}
\end{equation}
Its off-diagonal entry is $\cos^2\theta/2$, not $\cos(2\theta)/2$: it is the overlap of the two channel-conditioned spatial amplitudes and vanishes at $\theta=\pi/2$. Its eigenvalues are
$\lambda_\pm=(1\pm\cos^2\theta)/2$, and the entanglement entropy is
\begin{equation}
 S(t)=-\sum_{\sigma=\pm}\lambda_\sigma\log_2\lambda_\sigma.
 \label{eq:router_entropy}
\end{equation}
Thus $S(0)=0$ and $S(t_*)=1$.  Measuring the internal state in the basis
$|\pm\rangle=(|1_c\rangle\pm|2_c\rangle)/\sqrt2$ prepares the spatial states
$(|2\rangle\pm|4\rangle)/\sqrt2$, providing a simple path-entanglement/quantum-eraser interpretation.

\begin{theorem}[Entangling router]\label{thm:entangling_router}
Let a vertex $v$ have $d$ incident edges $(v,u_\alpha)$ with mutually orthogonal colours $\alpha=1,\ldots,d$, and suppose each edge Hamiltonian has the two-site form~(\ref{eq:caging_edge_hamiltonian}) with positive coupling strength $\kappa_\alpha$ for each edge, and all have a common perfect-transfer time $t_*$ (i.e. $\kappa_\alpha t_*/\hbar$ is an odd multiple of $\pi/2$).  Then
\begin{equation}
 |v\rangle\otimes\frac1{\sqrt d}\sum_{\alpha=1}^d|\alpha\rangle
 \quad\xrightarrow{\ t_*\ }\quad
 \frac1{\sqrt d}\sum_{\alpha=1}^d|u_\alpha\rangle|\alpha\rangle.
 \label{eq:general_router_map}
\end{equation}
The final state has Schmidt rank $d$, entropy $\log_2d$, and zero amplitude at every vertex outside $\{u_1,\ldots,u_d\}$.
\end{theorem}
\textbf{Proof}: In channel $\alpha$, proper colouring leaves only the matched edge $(v,u_\alpha)$ incident on $v$.  At $t_*$, eq.~(\ref{eq:edge_transfer_amplitudes}) maps $|v\rangle$ to $|u_\alpha\rangle$: for $\theta_\alpha=(2n_\alpha+1)\pi/2$, one has $i e^{-i\theta_\alpha}\sin\theta_\alpha=1$, independently of $n_\alpha$.  Orthogonality of the projectors prevents cross-channel terms.  The neighbours are distinct, so the displayed sum is already a Schmidt decomposition with equal coefficients. \hfill Q.E.D.

A physical realization requires distinguishable internal states at each vertex, such as optical modes or atomic internal levels. Hopping along an edge must preserve the corresponding internal state, while transitions between different states require additional controls. The edge-colouring bound determines the number of internal states sufficient for the construction, but does not constrain the number of physical couplers or their accuracy.

The controlled direct sum in eq.~(\ref{eq:controlled_hamiltonian}) is a standard controlled Hamiltonian~\cite{barenco}, and the resulting entanglement operation is conventional. The algebraic contribution is to select compatible matching-supported kinetic blocks; independently specified internal-state preparation and control supply the router. Exact confinement holds as long as the Hamiltonian preserves the channel decomposition.

\subsubsection{Exact invariant cages and finite-time leakage bounds}
Let $\Pi_{\rm cage}$ project onto a selected union of matching components and let $\Pi_{\rm cage}^\perp=1-\Pi_{\rm cage}$ denote its orthogonal complement.  If a perturbation $W_{\rm pert}$ preserves the idempotent sectors and the selected matching components, then
\begin{equation}
 \Pi_{\rm cage}^\perp H_0\Pi_{\rm cage}=\Pi_{\rm cage}^\perp W_{\rm pert}\Pi_{\rm cage}=0
 \quad\Longrightarrow\quad
 \Pi_{\rm cage}^\perp e^{-it(H_0+W_{\rm pert})/\hbar}\Pi_{\rm cage}=0
 \label{eq:exact_cage_protection}
\end{equation}
for every time and perturbation strength.  Therefore onsite disorder, edge-strength disorder and channel-dependent phases do not cause leakage provided they do not mix sectors or add edges outside the matching components.  For a small channel-mixing perturbation $H_\epsilon=H_0+\epsilon W_{\rm pert}$, Duhamel's formula gives the norm bound
\begin{equation}
 \left\|\Pi_{\rm cage}^\perp e^{-itH_\epsilon/\hbar}\Pi_{\rm cage}\right\|
 \leq \frac{|\epsilon|\,|t|}{\hbar}\|W_{\rm pert}\|,
 \qquad
 p_{\rm leak}(t)\leq
 \min\!\left\{1,\frac{\epsilon^2t^2}{\hbar^2}\|W_{\rm pert}\|^2\right\}.
 \label{eq:leakage_bound}
\end{equation}
The bound controls leakage perturbatively only when $|\epsilon t|\|W_{\rm pert}\|/\hbar$ is small; for later times the estimate becomes trivial and does not establish either a uniform-in-time protection theorem or a noise threshold. A stronger off-resonant estimate would require a separate spectral-gap assumption, which is not made here. For a synchronized two-site router, a timing error $\delta\theta$ gives infidelity $1-\sin^2(\pi/2+\delta\theta)=\sin^2\delta\theta=\delta\theta^2+O(\delta\theta^4)$.

\subsubsection{Arbitrary-graph realization and collective dark states}
\label{sec:general_matching_dark}
A subspace is called port-dark when it is Hamiltonian-invariant and lies in the kernel of the designated port-coupling map. Canonical compatibility admits a matching-layer realization on every finite simple graph; no converter, regularity, bipartiteness, or port hypothesis is required.  The existence and classification of a collective port-dark subspace additionally require a specified Hilbert-space restriction, a structured onsite converter, and a specified class of observation ports.  We first isolate the unconditional construction.

By the canonical compatibility criterion and Vizing's theorem, every finite simple graph admits a matching-based realization over $\mathbb C^q$, with $q\leq\Delta(G)+1$. Assigning a positive coupling strength $\gamma_e$ to each edge gives $w_e=\gamma_e\varepsilon_{c(e)}$, where $c$ is a proper edge colouring. The construction applies to arbitrary finite graphs, independently of their size, regularity or planarity. Positivity ensures that each weighted Laplacian channel is nonnegative; canonical compatibility itself does not require positivity.

Transitions between different channels require additional couplings that are not determined by the Lie-bialgebra construction. Interparticle interactions may also be introduced independently, including interactions that preserve the confinement of particles to their respective cages.

\subsubsection{Iterated heterogeneous decorations and bounded-channel realizations}
\label{sec:iterated_decorations}
The universal finite-graph statement extends to families constructed by successive decorations.  Such an extension concerns the matching architecture, not automatic preservation of dynamical observables.  In particular, independently recolouring each finite graph does not supply a compatible family of embedded physical channel spaces; the channel assignment must be fixed on previously present edges.

\begin{proposition}[Compatible colouring under iterated decorations]
\label{prop:iterated_decoration}
Let $G_0\subseteq G_1\subseteq\cdots$ be finite simple graphs whose vertex and edge sets are nested, with existing edges and their endpoint identifications retained.  The added finite motifs may be nonisomorphic.  Suppose there exists a positive integer $\Delta_{\max}$ such that every vertex of every $G_n$ has degree at most $\Delta_{\max}$. Then the graphs admit a common number $q\le\Delta_{\max}+1$ of channels and proper colourings $c_n:E(G_n)\to\{1,\ldots,q\}$ with $c_{n+1}|_{E(G_n)}=c_n$. Thus, when new edges are added, the colours of all existing edges remain unchanged. Assigning fixed, strictly positive real amplitudes $\gamma_e>0$ and
\begin{equation}
 w_e=\gamma_e\varepsilon_{c_n(e)}\in\mathbb C^q
 \label{eq:decoration_weight}
\end{equation}
produces a canonically compatible matching-layer realization at every stage.  No bipartiteness, periodicity, or common motif type is required.
\end{proposition}
\textbf{Proof.} The union $G_\infty=\bigcup_nG_n$ is a countable simple graph of maximum degree at most $\Delta_{\max}$.  A compactness argument applied to the line graph (the de Bruijn--Erdos colouring theorem~\cite{debruijn_erdos}) gives a proper colouring of $G_\infty$ with at most $\Delta_{\max}+1$ colours: for each finite initial edge set choose a subsequence of colourings that agrees there, and diagonalize.  Its restrictions are the $c_n$.  Distinct incident edges have different colours, whence $w_ew_f=0$ for distinct incident $e,f$.  The canonical compatibility criterion applies at every stage. \hfill Q.E.D.

Consider a finite graph $F$ with maximum vertex degree at most $\Delta_{\max}$. Attach a pendant vertex $p_x$ to each vertex $x$ of $F$. The new edges form a matching
$\mathcal M_{\rm pend}=\{xp_x:x\in V(F)\}$, which can be assigned a single channel. The original edges of $F$ require at most $\Delta_{\max}+1$ additional channels. Thus the decorated graph admits a colouring with at most $\Delta_{\max}+2$ channels.

In the pendant channel, each pair $(x,p_x)$ forms an independent two-site cage. For fermions, let $N_x=n_{x,*}+n_{p_x,*}$ denote the particle number in this cage. Consider the interaction
\begin{equation}
 H_{\mathrm{int}}(t)=\sum_{\{x,y\}\in E(F)}U_{xy}(t)n_{x,*}n_{y,*},
 \qquad [H_{\mathrm{int}}(t),N_x]=0
 \label{eq:decoration_density_interaction}
\end{equation}
The interaction preserves the particle number in every cage. In the sector $N_x=1$, the states $|x\rangle$ and $|p_x\rangle$ encode a qubit. Interactions between neighbouring cages produce conditional phases and, together with independently controlled single-cage operations, allow graph-state entanglement to be generated on $F$ \cite{hein_graphstates}.

The interactions and controls required for this construction are independent of the Lie-bialgebra axioms. Exact confinement is maintained when the complete Hamiltonian commutes with every $N_x$; additional hopping between cages need not satisfy this condition.

The preceding construction also illustrates how compatible colourings can be maintained as a graph grows. The colourings established by the proposition are chosen consistently for the entire sequence: new edges receive colours, while existing edges retain theirs. If new edges must instead be coloured as they arrive, without knowledge of subsequent extensions, a greedy procedure guarantees at most $2\Delta_{\max}-1$ colours for $\Delta_{\max}\geq2$. The smaller bound $\Delta_{\max}+1$ therefore applies when the full sequence is taken into account.

These bounds concern the number of channels, not the complexity of their physical implementation. The number and accuracy of the required couplers, as well as the control of transitions between channels, depend on the particular realization. Imperfections that do not preserve the relevant invariant subspaces may also destroy exact confinement or routing.

A further distinction concerns the Lie-bialgebra structures themselves. Although the graphs form a nested sequence, their Lie bialgebras need not do so. Adding an edge changes the canonical cobracket at its endpoints, and extending vertex functions by zero does not necessarily define a Lie-bialgebra embedding. The proposition establishes compatible colourings and canonical edge weights at every stage, rather than embeddings between successive Lie bialgebras.

Related constructions involving collective port-dark states are discussed in Appendix~\ref{app:dark}. They use the kernel of the unsigned incidence matrix $\mathsf U$, with unit entries, whose relation to the line-graph adjacency matrix,
$\mathsf U^*\mathsf U-2I=\mathsf A_{L(G)}$, is a standard result of graph spectral theory~\cite{biggs_graph}. Under the specified additional couplings, this kernel yields invariant dark subspaces of the channel-resolved Hamiltonian. The appendix includes the nonplanar $K_{3,3}$ example and the construction based on even cycles.

\subsection{Fokker--Planck dynamics and first-passage caging over an ordered zero-divisor algebra}
\label{subsec:FP_ordered_ring}

We now consider stochastic evolution on the same weighted graphs, using the real ordered algebra $R=\mathbb R^q$ introduced above. The idempotents, edge weights, zero-product condition, and nonvanishing edge squares are unchanged. As in the quantum construction, the algebra fixes the operators but not the physical dynamics: stochastic evolution additionally requires a probability-conserving generator with nonnegative transition rates.

\subsubsection{The equation}

With the weighted Laplacian $\Delta_w$ defined in eq.~(\ref{eq:weighted_graph_laplacian}), set
\begin{equation}
        L=-\frac12\Delta_w.
\end{equation}
By Corollary~\ref{cor:common_transport_operator},
\begin{equation}
 L=\frac12\operatorname{ad}_1=-\frac12\partial\delta,
 \qquad
 L^2=\frac12\delta^\dagger\delta.
 \label{eq:FP_factorization_recalled}
\end{equation}
These identities distinguish first and second differential order in the graph calculus, not two independently measurable transport mechanisms. When $A$ and $B$ are diagonal, incident annihilation restricts the off-diagonal entries of both contributions to the same matching edges; a nondiagonal $B$ need not preserve this support. Spatially varying departure-site coefficients may produce unequal forward and reverse rates, but algebraic differential order alone does not define continuum drift. The constitutive matrices are independent input data, not quantities fixed by the cobracket. With $\mathfrak d_i=\sum_jw_{ij}$ denoting the weighted degree, the matrix entries of $L$ are
\begin{equation}
        L_{ii}=-\frac12\mathfrak d_i,
        \qquad
        L_{ij}=\frac12 w_{ij}\quad (i\neq j).
\label{eq:L_entries_ordered}
\end{equation}
Since the graph is undirected and $w_{ij}=w_{ji}$, one has
\begin{equation}
        \sum_i (Lf)_i=0,
        \qquad
        \sum_i (L^2f)_i=0.
\label{eq:L_column_conservation}
\end{equation}
Equivalently,
\begin{equation}
        {\bf 1}^T L=0,
        \qquad
        {\bf 1}^T L^2=0.
\end{equation}

Let
\begin{equation}
        A=\mathrm{diag}(a_1,\ldots,a_N),
        \qquad a_i\in R,
\end{equation}
and let $B\in R^{N\times N}$. The graph Fokker--Planck equation is
\begin{equation}
        \frac{dP}{dt}
        =
        -L(AP)+L^2(BP)
        =\frac12\partial\delta(AP)
         +\frac12\delta^\dagger\delta(BP)
        =MP,
        \qquad
        M=-LA+L^2B.
\label{eq:FPE_ordered_ring}
\end{equation}
Here $P(t)\in R^N$. Since $R=\mathbb R^q$, this is a finite collection of real linear systems for $P_i^{(\alpha)}(t)$. We use the term graph Fokker--Planck equation for its first- and second-order algebraic form; stochastic admissibility concerns the complete matrix $M$, not its individual summands. In particular, $L^2=Q/2$ has nonpositive off-diagonal entries for nonnegative edge weights and is not itself a Markov generator. The two-vertex spectral description holds separately in each channel; averaging the ring-valued operator before squaring generally changes the result. The set of $R$-valued probability vectors is
\begin{equation}
        {\mathcal S}_R
        :=
        \left\{P\in R^N:
        P_i\in R_+\textrm{ for all }i,
        \quad
        \sum_i P_i=1_R
        \right\}.
\label{eq:R_probability_simplex}
\end{equation}
Since $R=\mathbb R^q$, the set ${\mathcal S}_R$ is naturally the Cartesian product of $q$ ordinary probability simplices, ${\mathcal S}_R\cong(\Delta_{N-1})^q$. Thus each component $(P_i^{(\alpha)})_i$ is an ordinary conditional probability vector. A physical joint law additionally requires a prior $\pi_\alpha\ge0$, $\sum_\alpha\pi_\alpha=1$, giving $p_{i,\alpha}=\pi_\alpha P_i^{(\alpha)}$. The position marginal is obtained from the corresponding positive normalized functional $\mathfrak m:R\to\mathbb R$; explicitly,
\begin{equation}
        p_i=\mathfrak m(P_i),
        \qquad
        \mathfrak m(1_R)=1,
        \qquad
        \mathfrak m(R_+)\subset {\mathbb R}_{\geq0}.
\end{equation}

\subsubsection{Probability conservation and current}

From (\ref{eq:L_column_conservation}) it follows immediately that
\begin{equation}
        {\bf 1}^T M
        =
        -{\bf 1}^TLA+{\bf 1}^TL^2B
        =0.
\label{eq:M_conservative}
\end{equation}
Therefore
\begin{equation}
        \frac{d}{dt}\sum_iP_i(t)
        =
        \sum_i\dot P_i(t)
        =
        {\bf 1}^TMP(t)
        =0.
\end{equation}
Hence the total $R$-valued mass is conserved independently of positivity. For $i\neq j$, define the one-way flux from $j$ to $i$ by
\begin{equation}
        F_{i\leftarrow j}=M_{ij}P_j,
\end{equation}
and define the oriented probability current by
\begin{equation}
        J_{ij}=F_{i\leftarrow j}-F_{j\leftarrow i}
        =M_{ij}P_j-M_{ji}P_i.
\label{eq:FP_current_ordered_ring}
\end{equation}
Then
\begin{equation}
        J_{ji}=-J_{ij}.
\end{equation}
Using (\ref{eq:M_conservative}), the diagonal entries satisfy
\begin{equation}
        M_{ii}=-\sum_{j\neq i}M_{ji},
\end{equation}
and hence
\begin{eqnarray}
        \dot P_i
        &=&M_{ii}P_i+\sum_{j\neq i}M_{ij}P_j                                      \nonumber\\
        &=&-\sum_{j\neq i}M_{ji}P_i+\sum_{j\neq i}M_{ij}P_j          
               \nonumber\\
        &=&\sum_{j\neq i}\left(M_{ij}P_j-M_{ji}P_i\right)                            \nonumber\\
        &=&\sum_{j\neq i}J_{ij}.
\label{eq:local_conservation_ordered_ring}
\end{eqnarray}
The current is antisymmetric and need not be nonnegative, even when the one-way fluxes belong to $R_+$. For diagonal $B$, these currents follow the original edges; for general nondiagonal $B$, they belong to the effective transition graph defined by the off-diagonal entries of $M$.

\subsubsection{Effect of the zero-product rule}

Under the local zero-product condition (\ref{eq:zero_product_incident}), the second-order operator $L^2$ simplifies. Let
\begin{equation}
        s_i^{(2)}=\sum_{r\sim i} w_{ir}^2.
\end{equation}
Then
\begin{equation}
        (L^2)_{ii}=\frac12s_i^{(2)},
        \label{eq:L2_diagonal_ordered}
\end{equation}
and, for $i\neq j$,
\begin{equation}
        (L^2)_{ij}= 
        \left\{
        \begin{array}{ll}
        -\frac12 w_{ij}^2, & i\sim j, \\
        0, & i\not\sim j,
        \end{array}
        \right.
\label{eq:L2_offdiag_ordered}
\end{equation}
Mixed two-edge terms vanish by the incident zero-product rule~(\ref{eq:zero_product_incident}), whereas the same-edge square survives by eq.~(\ref{eq:square_survives}).

Consequently, if $B=\mathrm{diag}(b_1,\ldots,b_N)$, then for adjacent
$i\neq j$
\begin{equation}
        M_{ij}
        =
        -\frac12 w_{ij}a_j
        -\frac12 w_{ij}^2b_j.
\label{eq:Mij_diagonal_B_ordered}
\end{equation}
For a general non-diagonal $B$, one obtains
\begin{equation}
        M_{ij}
        =
        -\frac12 w_{ij}a_j
        +\frac12 s_i^{(2)}B_{ij}
        -\frac12\sum_{r\sim i}w_{ir}^2B_{rj},
        \qquad i\neq j,
        \label{eq:Mij_general_B_ordered}
\end{equation}
where $w_{ij}=0$ if $i$ and $j$ are not adjacent. Thus nondiagonal constitutive couplings may generate effective transitions beyond the original edges. The negative terms in eq.~(\ref{eq:Mij_diagonal_B_ordered}) do not impose negative rates: $B$ is constitutive data, while the physical transition rates are the off-diagonal entries of $M$. For example, with $A=0$, choose
\begin{equation}
        b_j=-2\nu_j,
        \qquad \nu_j\in R_+.
\end{equation}
Then
\begin{equation}
        M_{ij}=w_{ij}^2\nu_j\in R_+,
        \qquad i\sim j,
        \label{eq:pure_diffusion_rates_positive}
\end{equation}
Thus the sign convention in eq.~(\ref{eq:FPE_ordered_ring}) is consistent with Markov positivity: the nonnegative departure coefficient is $\nu_j=-b_j/2$, not $b_j$.

\subsubsection{Positivity of the solution}

\begin{definition}
Following the usual finite-state convention \cite{anderson_ctmc}, a matrix $M\in R^{N\times N}$ is called an $R$-Markov generator if
\begin{equation}
        M_{ij}\in R_+,
        \qquad i\neq j,
        \label{eq:R_Markov_offdiag}
\end{equation}
and
\begin{equation}
        \sum_iM_{ij}=0
        \qquad\textrm{for every }j.
\label{eq:R_Markov_column}
\end{equation}
\end{definition}

\begin{theorem}
The set ${\mathcal S}_R$ in (\ref{eq:R_probability_simplex}) is forward
invariant under $\dot P=MP$ if and only if $M$ is an $R$-Markov generator.
\end{theorem}

\textbf{Proof.}
Write
\begin{equation}
        M_{ij}=\sum_\alpha M_{ij}^{(\alpha)}\varepsilon_\alpha,
        \qquad
        P_i=\sum_\alpha P_i^{(\alpha)}\varepsilon_\alpha.
\end{equation}
The equation $\dot P=MP$ is equivalent to the family of real equations
\begin{equation}
        \dot P^{(\alpha)}=M^{(\alpha)}P^{(\alpha)},
        \qquad \alpha=1,\ldots,q.
\end{equation}
Condition (\ref{eq:R_Markov_column}) says that every real matrix
$M^{(\alpha)}$ has zero column sums, and condition
(\ref{eq:R_Markov_offdiag}) says that every off-diagonal entry of every
$M^{(\alpha)}$ is nonnegative. Hence each $M^{(\alpha)}$ is an ordinary
finite-state continuous-time Markov generator, and it preserves the ordinary
probability simplex. Therefore ${\mathcal S}_R$ is forward invariant.
Conversely, if ${\mathcal S}_R$ is forward invariant, apply the same argument
componentwise at the boundary of each real simplex. The vector field must point
inward or tangent to every face, which forces
$M_{ij}^{(\alpha)}\geq0$ for $i\neq j$, while conservation of total mass
forces the column sums to vanish. This is the finite-dimensional Nagumo condition applied componentwise \cite{nagumo}.  Q.E.D. 

For eq.~(\ref{eq:FPE_ordered_ring}), column sums vanish automatically by eq.~(\ref{eq:M_conservative}); admissibility therefore reduces to
\begin{equation}
        \left(-LA+L^2B\right)_{ij}\in R_+,
        \qquad i\neq j.
\label{eq:admissible_B_condition}
\end{equation}
For fixed $A$, the admissible $B$ form a convex set defined by affine componentwise inequalities. The following proposition proves that this set is nonempty and realizes arbitrary directed rates on the matching support.

\begin{proposition}[Construction of Markov generators with prescribed rates]
\label{prop:constructive_markov_admissibility}
Let $G$ be a finite simple graph with a proper edge colouring
$c:E(G)\to\{1,\ldots,q\}$ and weights
$w_e=\gamma_e\varepsilon_{c(e)}\in R=\mathbb R^q$, where
$\gamma_e>0$.  Take any prescribed diagonal
$A=\operatorname{diag}(a_j)$, with $a_j\in R$.
No independent sign constraint is imposed on the constitutive matrix
$B$. For each directed end $j\to i$ of an edge $e=\{i,j\}$ of
colour $\alpha$, prescribe a rate $r_{j\to i}^{(\alpha)}\ge0$.
Then there exists a diagonal $B=\operatorname{diag}(b_j)$ such that
$M=-LA+L^2B$ is an $R$-Markov generator whose off-diagonal rates
are precisely the prescribed matching-supported rates, with no
additional off-matching transitions.  The rates in the two directions
of an edge may be prescribed independently.  A vertex--channel pair $(j,\alpha)$ is called active if $j$ is incident to an edge of colour $\alpha$, and inactive otherwise. Since the edge colouring is proper, each active pair determines a unique incident edge $e=\{i,j\}$ with $c(e)=\alpha$. For fixed $A$, the prescribed transition rate $r_{j\to i}^{(\alpha)}$ uniquely determines $b_j^{(\alpha)}$. For an inactive pair, $b_j^{(\alpha)}$ may be chosen arbitrarily, since it does not affect the generator $M$.
\end{proposition}
\textbf{Proof.} By proper colouring, every vertex $j$ is incident to
at most one edge $e=\{i,j\}$ of colour $\alpha$.  On that edge,
eq.~(\ref{eq:Mij_diagonal_B_ordered}) reads
\begin{equation}
 M_{ij}^{(\alpha)}=-\tfrac12\gamma_e a_j^{(\alpha)}
                       -\tfrac12\gamma_e^2 b_j^{(\alpha)}.
\label{eq:markov_active_rate}
\end{equation}
Choose
\begin{equation}
 b_j^{(\alpha)}=-\frac{a_j^{(\alpha)}}{\gamma_e}
                  -\frac{2r_{j\to i}^{(\alpha)}}{\gamma_e^2}.
\label{eq:constructive_admissible_B}
\end{equation}
Division is by the nonzero real scalar $\gamma_e$, not by the
zero divisor $w_e$. Substitution into~(\ref{eq:markov_active_rate})
gives $M_{ij}^{(\alpha)}=r_{j\to i}^{(\alpha)}$ exactly.
Different directed rates involve different vertex--channel
components because each colour class is a matching; the reverse
rate uses $b_i^{(\alpha)}$.  For a vertex with no incident edge of
colour $\alpha$, the corresponding columns of $L^{(\alpha)}$ and
$(L^{(\alpha)})^2$ vanish, leaving $b_j^{(\alpha)}$ arbitrary.
All off-matching off-diagonal entries of $M$ vanish for diagonal
$B$, by~(\ref{eq:L2_offdiag_ordered}).  Finally,
$\mathbf 1^TM=0$ fixes each diagonal entry as the negative sum of
outgoing rates. Hence $M$ is componentwise Markov-admissible.
\hfill Q.E.D.

The proposition realizes every prescribed matching-supported rate assignment within the diagonal constitutive class.

\subsubsection{Gibbs-type reversible stationary states}

Let $\mathscr V_i\in R$ be a graph potential, with componentwise exponential
\begin{equation}
        \exp(-\mathscr V_i)
        =
        \sum_\alpha \exp\left(-\mathscr V_i^{(\alpha)}\right)\varepsilon_\alpha.
\end{equation}
Assume
\begin{equation}
        Z=\sum_i\exp(-\mathscr V_i)
\end{equation}
is strictly positive, i.e. every component of $Z$ is positive. Then $Z$ is invertible in $R$, and
\begin{equation}
        \rho_i^G=Z^{-1}\exp(-\mathscr V_i)
        \label{eq:ring_gaussian_state}
\end{equation}
defines a strictly positive $R$-valued probability vector. It is a finite-state Gibbs-type law, not a Gaussian distribution. To construct reversible dynamics, take $A=0$ and choose 
\begin{equation}
        B=-2c\,\mathrm{diag}\left((\rho_1^G)^{-1},\ldots,(
        \rho_N^G)^{-1}\right),
        \qquad c\in R_+.
\label{eq:B_for_gaussian}
\end{equation}
Using (\ref{eq:L2_offdiag_ordered}), the off-diagonal rates are
\begin{equation}
        M_{ij}
        =
        c\,w_{ij}^2(\rho_j^G)^{-1},
        \qquad i\sim j.
\label{eq:gaussian_rates}
\end{equation}
They are nonnegative elements of $R$. Moreover,
\begin{equation}
        M_{ij}\rho_j^G
        =
        c\,w_{ij}^2
        =
        M_{ji}\rho_i^G.
\label{eq:detailed_balance_gaussian}
\end{equation}
Hence $\rho^G$ is stationary and satisfies detailed balance. The associated current is
\begin{equation}
        J_{ij}
        =
        c\,w_{ij}^2
        \left(
        \frac{P_j}{\rho_j^G}
        -
        \frac{P_i}{\rho_i^G}
        \right).
\label{eq:gaussian_current_ordered}
\end{equation}
The current vanishes at $P=\rho^G$ and, more generally, whenever $P_i/\rho_i^G=P_j/\rho_j^G$ on the edge. It need not vanish away from equilibrium.

\subsubsection{Detailed balance and the H theorem}

Detailed balance also yields the standard entropy-dissipation law componentwise~\cite{maas_entropy}. Because each channel graph is a matching, its connected-component masses are conserved separately; a positive stationary state is therefore not generally unique. Let $\rho\in\mathcal S_R$ have strictly positive entries. We say that an $R$-Markov generator $M$ satisfies detailed balance with respect to $\rho$ if
\begin{equation}
        M_{ij}\rho_j=M_{ji}\rho_i=:W_{ij},
        \qquad i\neq j,
        \label{eq:detailed_balance_ordered}
\end{equation}
with
\begin{equation}
        W_{ij}=W_{ji}\in R_+.
\end{equation}
Then $M\rho=0$, and
\begin{equation}
        J_{ij}
        =
        W_{ij}
        \left(
        \frac{P_j}{\rho_j}
        -
        \frac{P_i}{\rho_i}
        \right).
\label{eq:current_detailed_balance_ordered}
\end{equation}
The nonnegative quantity entering the H theorem is the equilibrium conductance $W_{ij}$, not the raw weight $w_{ij}$. For strictly positive $P_i$, define the componentwise relative entropy
\begin{equation}
        {\mathcal D}_R(P\parallel\rho)
        =
        \sum_i P_i\log\left(P_i\rho_i^{-1}\right)
        \in R,
        \label{eq:KL_R}
\end{equation}
where logarithm is taken componentwise in $R={\mathbb R}^q$. Componentwise,
(\ref{eq:KL_R}) is exactly the ordinary Kullback--Leibler divergence.

\begin{theorem}
If $M$ is an $R$-Markov generator satisfying detailed balance with respect
to a strictly positive $\rho$, then along every strictly positive solution of
$\dot P=MP$
\begin{equation}
        \frac{d}{dt}{\mathcal D}_R(P\parallel\rho)\leq 0
\end{equation}
in the order of $R$.
\end{theorem}

\textbf{Proof.}
Set
\begin{equation}
        x_i=P_i\rho_i^{-1}.
\end{equation} 
Using conservation of total mass, local conservation, and detailed balance, the same symmetrisation as in the real finite-state Markov case gives
\begin{equation}
        \frac{d}{dt}{\mathcal D}_R(P\parallel\rho)
        =
        -\frac12\sum_{i,j}
        W_{ij}(x_j-x_i)(\log x_j-\log x_i).
\label{eq:KL_dissipation_R}
\end{equation}
For each component $\alpha$, the real inequality
\begin{equation}
        (x-y)(\log x-\log y)\geq0
        \qquad (x,y>0)
\end{equation}
implies that the $\alpha$-component of every term on the right-hand side of
(\ref{eq:KL_dissipation_R}) is nonpositive, because $W_{ij}^{(\alpha)}\geq0$. Hence the derivative belongs to $-R_+$.  This is precisely
\begin{equation}
        \frac{d}{dt}{\mathcal D}_R(P\parallel\rho)\leq0.
\end{equation}
Q.E.D.

Equality in eq.~(\ref{eq:KL_dissipation_R}) means that $P_i^{(\alpha)}/\rho_i^{(\alpha)}$ is constant on each connected component of the positive-rate graph in channel $\alpha$; disconnected matching components therefore need not relax to $\rho$ globally. A logarithm-free quadratic functional is
\begin{equation}
        {\mathcal E}_R(P\mid\rho)
        =
        \frac12\sum_i \rho_i\left(\frac{P_i}{\rho_i}-1_R\right)^2.
\end{equation}
Then detailed balance gives
\begin{equation}
        \frac{d}{dt}{\mathcal E}_R(P\mid\rho)
        =
        -\frac12\sum_{i,j}
        W_{ij}\left(\frac{P_i}{\rho_i}-\frac{P_j}{\rho_j}\right)^2.
\label{eq:quadratic_entropy_R}
\end{equation}
Every summand on the right is nonnegative in $R$, so this gives the corresponding sum-of-squares dissipation identity.

\subsubsection{First-passage caging and stochastic routing}
\label{subsec:first_passage_caging}

We next use matching-supported rates to obtain exact first-passage exclusion and channel-dependent routing. These conclusions need not hold for a nondiagonal $B$ that introduces additional effective transitions. Write
\begin{equation}
 P_i(t)=\sum_{\alpha=1}^{q}P_i^{(\alpha)}(t)\varepsilon_\alpha
\end{equation}
and choose a prior distribution $\pi_\alpha$ for a classical internal channel $C$.  The ordinary joint probability is
\begin{equation}
 p_{i,\alpha}(t)=\pi_\alpha P_i^{(\alpha)}(t),
 \qquad
 \sum_{i,\alpha}p_{i,\alpha}(t)=1.
 \label{eq:joint_internal_probability}
\end{equation}
With the label conserved, the generator is block diagonal in $\alpha$, paralleling the channel-preserving Hamiltonian~(\ref{eq:controlled_hamiltonian}). First-passage observables detect this decomposition directly~\cite{redner,benichou,grebenkov2019}.

\begin{theorem}[Stochastic caging]\label{thm:stochastic_caging}
Consider the pure-diffusion family $A=0$ with coefficients
\[
 B=-2\operatorname{diag}(\nu_1,\ldots,\nu_N),\qquad \nu_i\in R_+.
\]
For a fixed channel $\alpha$, the effective transition graph of $M^{(\alpha)}$ is contained in the colour-$\alpha$ matching.  Hence a process initially at a vertex $v$ remains in the matching component of $v$, which contains at most two vertices.  If a target set $T$ is disjoint from that component, then
\begin{equation}
 \mathbb P_{v,\alpha}(\tau_T<\infty)=0,
 \qquad
 \mathbb E_{v,\alpha}[\tau_T]=\infty.
 \label{eq:stochastic_cage_fpt}
\end{equation}
\end{theorem}
\textbf{Proof}: By eq.~(\ref{eq:pure_diffusion_rates_positive}),
$M_{ij}^{(\alpha)}=(w_{ij}^2\nu_j)^{(\alpha)}$ can be nonzero only when the edge $ij$ has colour $\alpha$.  A proper edge colouring makes every colour class a matching, so no channel-$\alpha$ path can leave the isolated vertex or matched pair containing $v$.  A disjoint target is therefore unreachable. \hfill Q.E.D.

The same result gives an exact stochastic router.  Let a vertex $v$ have incident edges $(v,u_\alpha)$ of distinct colours $\alpha=1,\ldots,d$, with
\begin{equation}
 w_{vu_\alpha}=\gamma_\alpha\varepsilon_\alpha.
\end{equation}
Choose
\begin{equation}
 \nu_v=\sum_{\alpha=1}^{d}\frac{k_\alpha}{\gamma_\alpha^2}\varepsilon_\alpha,
 \qquad
 \nu_{u_\alpha}=0,
 \label{eq:stochastic_router_diffusion}
\end{equation}
and set all remaining departure-rate coefficients to zero, with $k_\alpha>0$ and $\gamma_\alpha\ne0$. The outputs are absorbing because their departure-rate coefficients vanish. Let $\tau$ be the first hitting time of the output set $U=\{u_1,\ldots,u_d\}$.  Then the only outgoing transition in channel $\alpha$ is $v\to u_\alpha$, at rate $k_\alpha$.  Conditional on $C=\alpha$,
\begin{equation}
 P_v^{(\alpha)}(t)=e^{-k_\alpha t},
 \qquad
 P_{u_\alpha}^{(\alpha)}(t)=1-e^{-k_\alpha t},
 \qquad
 P_{u_\beta}^{(\alpha)}(t)=0\quad(\beta\neq\alpha).
 \label{eq:stochastic_router_solution}
\end{equation}
Thus the splitting probabilities are
\begin{equation}
 \boxed{\mathbb P(X_{\tau}=u_\beta\mid C=\alpha)=\delta_{\alpha\beta}.}
 \label{eq:perfect_stochastic_splitting}
\end{equation}
Thus the conserved channel label determines the absorbing output exactly: the construction acts as an ideal stochastic demultiplexer.

For equal rates $k_\alpha=k$, the joint distribution in eq.~(\ref{eq:joint_internal_probability}) satisfies
\begin{equation}
 p_{v,\alpha}(t)=\pi_\alpha e^{-kt},
 \qquad
 p_{u_\alpha,\alpha}(t)=\pi_\alpha(1-e^{-kt}).
\end{equation}
If $\mathsf H(\pi)=-\sum_\alpha\pi_\alpha\log_2\pi_\alpha$, then observing the source vertex leaves the prior unchanged, whereas observing $u_\alpha$ determines the internal label exactly.  Therefore
\begin{equation}
 \boxed{I(C;X_t)=(1-e^{-kt})\mathsf H(\pi).}
 \label{eq:stochastic_mutual_information}
\end{equation}
For a uniform prior, the limiting mutual information is $\log_2d$, matching the numerical entanglement bound of the coherent router in Theorem~\ref{thm:entangling_router}, without identifying the two correlation measures.

\subsubsection{Escape from caging under weak channel switching}
\label{subsec:switching_escape}

To quantify controlled escape from exact caging, introduce onsite switching on the joint state space $V\times\{1,2\}$ for the coloured square in eq.~(\ref{eq:caging_square_weights}).  With column probability vectors, let $M^{(\alpha)}$ denote the pure-diffusion generator in channel $\alpha$, and define
\begin{equation}
 \mathcal G_\epsilon=
 \begin{pmatrix}M^{(1)}&0\\0&M^{(2)}\end{pmatrix}
 +\epsilon\begin{pmatrix}-I&I\\I&-I\end{pmatrix},
 \label{eq:enlarged_switching_generator}
\end{equation}
with transitions stopped on first arrival at vertex $3$ when computing hitting times. The switching term is not $R$-linear: it mixes the idempotent channels and is not implied by the original cobracket. The resulting process $(X_t,C_t)$ has spatial jumps at rate $k>0$ along the matching selected by $C_t$, and internal switches at rate $\epsilon>0$ at each nonabsorbing vertex. Let vertex $3$ be absorbing and denote by $T_{i,\alpha}$ the mean hitting time of vertex $3$ from $(i,\alpha)$. For $i\neq3$ the backward equations are
\begin{equation}
 -1=k\bigl(T_{j(i,\alpha),\alpha}-T_{i,\alpha}\bigr)
 +\epsilon\bigl(T_{i,\bar\alpha}-T_{i,\alpha}\bigr),
 \qquad T_{3,\alpha}=0,
 \label{eq:backward_switching_square}
\end{equation}
where
\begin{equation}
\begin{array}{lll}
 j(1,1)=2, & j(2,1)=1, & j(4,1)=3,\\
 j(1,2)=4, & j(4,2)=1, & j(2,2)=3.
\end{array}
\end{equation}
Solving this six-dimensional linear system gives
\begin{equation}
\begin{aligned}
 T_{1,1}=T_{1,2}&=\frac{4}{k}+\frac{2}{\epsilon},\\
 T_{2,1}=T_{4,2}&=\frac{3}{k}+\frac{2}{\epsilon},\\
 T_{4,1}=T_{2,2}&=\frac{3}{k}.
\end{aligned}
\label{eq:switching_square_mfpt_solution}
\end{equation}
In particular,
\begin{equation}
 \boxed{\mathbb E_{1,\alpha}[\tau_3]
 =\frac{4}{k}+\frac{2}{\epsilon}
 \sim\frac{2}{\epsilon}\quad(\epsilon\downarrow0).}
 \label{eq:singular_escape_law}
\end{equation}
At $\epsilon=0$ the target is inaccessible by Theorem~\ref{thm:stochastic_caging}; it becomes reachable for every $\epsilon>0$, with a mean escape time diverging as switching vanishes. This is a property of the enlarged Markov model, not of the channel-conserving $R$-linear equation.

A complementary two-output experiment makes vertices $2$ and $4$ absorbing and starts the process at $(1,1)$.  The two transient internal states at vertex $1$ switch at rate $\epsilon$ and exit at rate $k$ to the channel-selected output.  The backward equations for the probability of reaching the correct output $2$ give
\begin{equation}
 \boxed{P_{\rm correct}=\frac{k+\epsilon}{k+2\epsilon},
 \qquad
 P_{\rm error}=\frac{\epsilon}{k+2\epsilon}.}
 \label{eq:stochastic_router_error}
\end{equation}
Thus weak channel noise produces a linear routing error $P_{\rm error}=\epsilon/k+O(\epsilon^2)$, while the channel-conserving router remains exact.

\begin{theorem}[Nonuniform first passage on coloured paths]
\label{thm:nonuniform_coloured_path_fpt}
Let $G_\ell$ be the path on vertices $0,1,\ldots,\ell$, with
$\ell\geq2$, and let $c_e$ denote the colour of the edge
$e=\{e,e+1\}$, where $c_{e-1}\neq c_e$ for adjacent edges.
On the joint position--channel state space, prescribe symmetric transitions
$(e,c_e)\leftrightarrow(e+1,c_e)$ at rates $k_e>0$ and, independently,
onsite channel switches $(v,c_{v-1})\leftrightarrow(v,c_v)$ at rates
$\eta_v>0$ for $1\leq v\leq\ell-1$. No other transitions are permitted.
Starting at $(0,c_0)$, stop at the first arrival at
$(\ell,c_{\ell-1})$. Then the mean first-passage time is
\begin{equation}
 \boxed{\mathbb E\tau_\ell=\sum_{e=0}^{\ell-1}\frac{2e+1}{k_e}
 +\sum_{v=1}^{\ell-1}\frac{2v}{\eta_v}.}
 \label{eq:arbitrary_coloured_path_fpt}
\end{equation}
For uniform edge-transition and switching rates, $k_e=k$ and
$\eta_v=\eta$, this reduces to
\begin{equation}
 \boxed{\mathbb E\tau_\ell=\frac{\ell^2}{k}
 +\frac{\ell(\ell-1)}{\eta}.}
 \label{eq:uniform_coloured_path_fpt}
\end{equation}
\end{theorem}
\textbf{Proof.} In their linear order, the accessible states are
$(0,c_0),(1,c_0),(1,c_1),(2,c_1),\ldots,(\ell,c_{\ell-1})$.
They form a reversible birth--death chain of $2\ell$ states with
successive symmetric rates $r_{2e}=k_e$ ($0\leq e\leq\ell-1$)
and $r_{2v-1}=\eta_v$ ($1\leq v\leq\ell-1$).
With the last state absorbing, let $T_j$ be the mean hitting time from
state $j$, and write $\zeta_j=T_j-T_{j+1}$.
The backward equations give $r_0\zeta_0=1$ and
$r_j\zeta_j=1+r_{j-1}\zeta_{j-1}$, hence
$\zeta_j=(j+1)/r_j$ by induction.
Summing $T_0=\sum_{j=0}^{2\ell-2}\zeta_j$ yields
(\ref{eq:arbitrary_coloured_path_fpt}); the sums of odd and even
integers yield (\ref{eq:uniform_coloured_path_fpt}). \hfill Q.E.D.

For example, a path of six edges and seven vertices with the periodic
three-colour assignment $c_e=1+(e\bmod3)$ has, at uniform rates,
$\mathbb E\tau_6=36/k+30/\eta$. Two alternating colours also satisfy
the theorem. Each transition along an edge preserves its channel;
progress across successive differently coloured edges is possible only
because the onsite switches are prescribed separately. Thus the theorem
concerns an explicitly enlarged Markov process, not transport generated
by the channel-conserving $R$-linear operator alone.

\subsubsection{Interpretation}

The stochastic and quantum applications share the algebraic operator $Q=\delta^\dagger\delta$, but not a physical time-evolution generator. Here $M=-LA+L^2B$ is stochastic only under the Markov inequalities, and the matching-supported choice is $M=-Q\operatorname{diag}(\nu_v)$, not $+Q$. The realizability proposition supplies arbitrary nonnegative directed rates on the coloured matching edges for a fixed diagonal $A$. Detailed balance gives the standard entropy-dissipation law, while conserved channels yield exact caging and absorbing sorting. The switching examples introduce independent Markov controls and quantify first-passage escape, routing errors, and first-passage times on longer coloured paths. These distinctions preserve the common algebraic structure without conflating coherent and stochastic dynamics.

\section{Discussion}

The central result concerns compatibility with a specified same-weight edge-difference cobracket (Theorem~\ref{thm:cocycle}). The graph bracket alone obeys Jacobi's identity over an integral domain only when its nonzero support consists of uniformly weighted clique components. Compatibility with the canonical cobracket imposes the stronger condition that weights on distinct incident edges annihilate one another. Over finite products of integral domains, the result is a matching support in each coefficient component (Corollary~\ref{cor:complete_matching_layers}). This matching interpretation does not extend to arbitrary nonreduced rings, as the dual-number example in Remark~\ref{ex:dual_number} demonstrates.

The factorization $\Delta_w^2=2\delta^\dagger\delta$ follows from the complete weighted graph operator, with $\mathfrak d_i=\sum_jw_{ij}$ including all edges incident to each vertex. The annihilation of distinct incident edge weights cancels the mixed terms. Resolving the coefficient algebra into orthogonal idempotent channels then gives the spectral interpretation: each active edge contributes an independent two-vertex block. The factorization supplies the common edge operator underlying both physical realizations, while the channelwise matching structure determines their shared confinement geometry. Channelwise products are taken before any scalar averaging.

The two physical realizations use this structure in different ways. Orthogonal quantum internal-state projectors yield exact caging and fixed-channel two-site transfer, while coherent channel-selective controls enable path--channel entanglement. In the stochastic realization, matching-supported Markov rates give first-passage exclusion and absorbing sorting. Onsite channel switching connects the previously isolated sectors, with escape times and routing errors given by eqs.~(\ref{eq:singular_escape_law}) and (\ref{eq:stochastic_router_error}); the coloured-path theorem gives exact nonuniform first-passage times for the corresponding controlled dynamics. The scalar $C_4$ Hamiltonian serves as an external comparator rather than an instance of the scalar graph Lie bracket. The quantum generator is Hermitian, while the stochastic generator satisfies the conservation and positivity conditions for a Markov process. Thus the same edge calculus supports two distinct, well-defined dynamical constructions.

The algebraic matching structure provides a natural framework for the addition of physical controls. Otherwise isolated sectors are connected by channel converters, inter-cage density interactions permit conditional operations, and transfer protocols combine these ingredients for the chosen task. Proper edge colouring bounds the number of internal channel labels needed on graphs of bounded degree; physical coupler counts, fidelities and protocol costs are determined by the implementation. The appendix develops further spectral examples under stated converter, port and coupling assumptions. In particular, the $K_{3,3}$ cycle modes have port-dependent visibility and the specified perturbative protection, while the pendant-cage fermionic construction gives a conditional interacting example. These results illustrate concrete uses of the formalism without asserting a general classification of interacting systems.

The construction has a specific algebraic focus and complements existing graph formalisms. Discrete graph differential calculi \cite{dim,majid} address other aspects of graph geometry; here the vertex bracket admits an explicit clique classification and a canonical matching condition. The graph Lie bialgebras of Farinati--Jancsa \cite{farinati} arise from two-step nilpotent graph Lie algebras, whereas the idempotent matching factors here are generally nonnilpotent. The dual-number example has the quotient-type relation described in Remark~\ref{ex:dual_number}; its Killing form is singular. The contribution of the present construction is a compatible vertex--edge calculus with an exact operator factorization and a transparent channel structure that can be used with standard quantum controls and Markov dynamics.

The precise hypotheses also indicate how the formalism can be developed further. The equivalence theorem concerns the specified same-weight cobracket, with the zero cobracket providing an elementary alternative. Its converse uses the absence of nonzero $2$-torsion, and the matching-layer decomposition applies to finite products of integral domains. Nonreduced rings behave differently: they may accommodate arbitrary graph support with vanishing formal edge squares, and general zero-divisor rings need not decompose into matching layers. Noncommutative coefficients would call for a modified algebraic setting. For the stated controlled single-particle and finite-state systems, the present quantum and stochastic results establish exact statements, providing a starting point for spatially varying switching, resetting and more general interacting dynamics.

\section{Conclusions}

For the graph bracket and specified same-weight edge-difference cobracket, Lie-bialgebra compatibility is equivalent, under the stated ring assumptions, to the annihilation of weights on distinct incident edges. Over finite products of integral domains, this sharpens the clique classification obtained from Jacobi's identity to a matching support in each coefficient component. Orthogonal idempotents provide such realizations on arbitrary finite graphs. Their Lie algebras consist of affine edge factors and central channel-isolated directions; they are metabelian and generally nonnilpotent. The identities $\Delta_w=\partial\delta$ and $\Delta_w^2=2\delta^\dagger\delta$ provide the common operator structure underlying the quantum and stochastic applications.

The channel structure supports exact quantum caging and synchronized entangling routing when combined with the specified coherent controls; a uniform degree-$d$ router produces path--channel entanglement of entropy $\log_2d$. With nonnegative Markov rates, the stochastic construction gives channel-resolved first-passage exclusion and sorting, with mutual information approaching $\log_2d$. Prescribed channel switching connects the sectors and yields quantitative escape and error laws, including the coloured-square mean first-passage time $4/k+2/\epsilon$, the routing-error probability $\epsilon/(k+2\epsilon)$, and the exact nonuniform coloured-path first-passage formula. The formalism thus combines an algebraically determined channel structure with independently chosen dynamical controls. It provides exact, physically interpretable realizations while leaving room for further choices of couplings, interactions and protocols.

\appendix
\section{Conditional active-channel dark subspaces and decorations}\label{app:dark}
We now impose additional conditions for a separate spectral statement.  Assume $G$ is connected and has at least one edge; use the single-idempotent edge-colouring construction described in Section~\ref{sec:general_matching_dark} and take strictly positive real edge parameters $\gamma_e$.  The spectral classification below is for the resulting active edge-end system, not automatically for every physical choice of onsite controls on the full $|V|q$-dimensional channel space.

Write $d_v$ for the degree of $v$ and $I_v=\{c(e):e\ni v\}$ for the active channel labels at $v$. Proper colouring makes $|I_v|=d_v$.  We work in the invariant active-edge-end subspace
\begin{equation}
 \mathscr H_E=\bigoplus_{v\in V}\mathbb C^{I_v}\cong\mathbb C^{\{(v,e):v\in e\}},\qquad \dim\mathscr H_E=2|E|.
 \label{eq:general_active_end_space}
\end{equation}
It is a subspace of the full physical $\mathbb C^V\otimes\mathbb C^q$ realization.  On irregular graphs, unused channel states at a vertex are excluded from this specified invariant subspace; to realize the following theorem physically, onsite controls must not couple active and inactive channels.  For each edge $e=\{u,v\}$, let $J_e=\hbar^2\gamma_e^2/m>0$.  The squared-Laplacian kinetic Hamiltonian acts on the corresponding two endpoint states as $J_e\left(\begin{smallmatrix}1&-1\\-1&1\end{smallmatrix}\right)$.  Define the unsigned endpoint-summation map
\begin{equation}
 \mathsf B_{\rm port}:\mathscr H_E\longrightarrow\mathbb C^V,\qquad (\mathsf B_{\rm port}\psi)_v=\sum_{e\ni v}\psi_{v,e}.
 \label{eq:general_port_map}
\end{equation}
\textbf{Additional converter hypothesis.}  At every vertex $v$, require a Hermitian active-channel block of the form
\begin{equation}
 C_v=a I_{d_v}+b_v\mathbf1_v\mathbf1_v^\dagger,\qquad a,b_v\in\mathbb R,
 \label{eq:general_converter_family}
\end{equation}
with a common $a$ at all vertices, but otherwise arbitrary $b_v$.  For this to define a physical invariant subsystem on an irregular graph, require the complete onsite converter to preserve the active/inactive channel splitting; its inactive-channel block is otherwise unrestricted.  Denote $H=H_0+\bigoplus_v C_v$ on $\mathscr H_E$.  Darkness is defined relative to the active-channel-symmetric vertex ports $d_v^{-1/2}\sum_{e\ni v}|v,e\rangle$; equivalently it means annihilation by $\mathsf B_{\rm port}$.  It is not darkness to arbitrary channel-resolved probes.

Let $\mathsf U:\mathbb C^E\to\mathbb C^V$ be the unsigned vertex--edge incidence matrix, $(\mathsf Uz)_v=\sum_{e\ni v}z_e$, and choose an orientation on $E$ with oriented incidence matrix $\mathsf D_{\rm or}$.  Define $(\iota_+z)_{v,e}=z_e$ at both ends and $(\iota_-z)_{u,e}=z_e$, $(\iota_-z)_{v,e}=-z_e$ for the chosen orientation $e:u\to v$.  For each distinct kinetic strength $J>0$, let $F_J$ denote the subgraph with edge set
$E(F_J)=\{e\in E:J_e=J\}$ and vertex set consisting of its edge endpoints.
Write $c(F_J)$ for its number of connected components. Its first Betti number,
or cycle rank, is
$\beta_1(F_J)=|E(F_J)|-|V(F_J)|+c(F_J)$, the dimension of its cycle space.

\begin{theorem}[Conditional active-channel dark-sector classification]
\label{thm:general_matching_dark}
For the edge-colouring construction described in Section~\ref{sec:general_matching_dark}, assume (i) $G$ is connected with at least one edge; (ii) the dynamics are restricted to the invariant active-edge-end space~(\ref{eq:general_active_end_space}), with no active--inactive conversion; (iii) $J_e>0$ are real; (iv) the Hermitian active converters satisfy~(\ref{eq:general_converter_family}) with the same real $a$ at every vertex; and (v) darkness is defined relative only to the active-channel-symmetric vertex ports, equivalently $\ker\mathsf B_{\rm port}$.  Then the maximal $H$-invariant subspace in $\ker\mathsf B_{\rm port}$ has the orthogonal decomposition
\begin{equation}
 \mathscr D_E(H,\mathsf B_{\rm port})=\iota_+(\ker\mathsf U)\ \oplus\!\bigoplus_{J}\iota_-\big(\ker(\mathsf D_{\rm or}|_{E(F_J)})\big).
 \label{eq:general_dark_decomposition}
\end{equation}
The first summand has eigenvalue $a$; the summand indexed by $J$ has eigenvalue $a+2J$.  Consequently
\begin{equation}
 \dim\mathscr D_E(H,\mathsf B_{\rm port})=\dim\ker\mathsf U+\sum_J\beta_1(F_J).
 \label{eq:general_dark_count}
\end{equation}
Writing $n=|V|$ and $m=|E|$, for a connected graph
\begin{equation}
 \dim\ker\mathsf U=\begin{cases}m-n+1,&G\text{ bipartite},\\m-n,&G\text{ nonbipartite}.\end{cases}
 \label{eq:general_unsigned_rank}
\end{equation}
In particular, all states in $\iota_+(\ker\mathsf U)$ are invariant under arbitrary positive, even time-dependent, kinetic edge strengths $J_e(t)$ and converters of the form~(\ref{eq:general_converter_family}) with common $a(t)$; their evolution is multiplication by $\exp[-(i/\hbar)\int a(s)\,ds]$.  When $G$ is $q$-regular with a proper $q$-edge colouring, $\mathscr H_E=\mathbb C^V\otimes\mathbb C^q$ and the classification applies directly to the full physical Hilbert space.
\end{theorem}
\textbf{Proof.} The converters on $\mathscr H_E$ have the combined form
$C=aI+\mathsf B_{\rm port}^{\dagger}\operatorname{diag}(b_v)\mathsf B_{\rm port}$.
Hence $H\psi=(H_0+aI)\psi$ for every $\psi\in\ker\mathsf B_{\rm port}$.
A subspace of this kernel is $H$-invariant precisely when it is $H_0$-invariant.
By Hermiticity, the maximal such subspace is the direct sum of the intersections
of the eigenspaces of $H_0$ with $\ker\mathsf B_{\rm port}$.

On an edge $e=\{u,v\}$ the kinetic block is
$J_e\left(\begin{smallmatrix}1&-1\\-1&1\end{smallmatrix}\right)$.
Its symmetric endpoint vector has eigenvalue $0$, while its antisymmetric
endpoint vector has eigenvalue $2J_e$. Thus the zero-energy eigenspace is
$\iota_+(\mathbb C^E)$; for each distinct $J>0$, the eigenspace of energy
$2J$ consists exactly of antisymmetric endpoint vectors supported on $E(F_J)$.
These mutually orthogonal eigenspaces exhaust $\mathscr H_E$.

The port map satisfies the incidence identities
\begin{equation}
 \mathsf B_{\rm port}\iota_+z=\mathsf Uz,\qquad
 \mathsf B_{\rm port}\iota_-z=\mathsf D_{\rm or}z,
 \label{eq:dark_incidence_port_identities}
\end{equation}
where the second identity uses the fixed orientation. Intersecting each kinetic
eigenspace with the port kernel gives exactly the direct sum in
(\ref{eq:general_dark_decomposition}), with energies $a$ and $a+2J$.
The oriented incidence matrix of $F_J$ has rank $|V(F_J)|-c(F_J)$, so
$\dim\ker(\mathsf D_{\rm or}|_{E(F_J)})=\beta_1(F_J)$.
For the unsigned incidence matrix, $\mathsf U^\dagger y=0$ means
$y_u+y_v=0$ on each edge. Connectivity leaves one scalar degree of freedom
for a bipartite graph; an odd cycle forces it to vanish. Rank--nullity yields
(\ref{eq:general_unsigned_rank}) and the complete dimension formula
(\ref{eq:general_dark_count}).

Finally, for $z\in\ker\mathsf U$, every symmetric edge-end vector
$\iota_+z$ is annihilated edge by edge by $H_0(t)$, regardless of the
positive strengths $J_e(t)$. Equation~(\ref{eq:dark_incidence_port_identities})
also shows that its converter contribution vanishes apart from the common
scalar $a(t)$. Therefore $H(t)\iota_+z=a(t)\iota_+z$, and the exact
propagator on this sector is multiplication by
$\exp[-(i/\hbar)\int_0^t a(s)\,ds]$. \hfill Q.E.D.

For a connected bipartite graph the symmetric protected sector has dimension $\beta_1(G)$; for a connected nonbipartite graph it has dimension $\beta_1(G)-1$. Additional antisymmetric dark modes are determined by the equal-strength subgraphs $F_J$. The unsigned-incidence kernel is the $-2$ adjacency eigenspace of the line graph, since $A(L(G))=\mathsf U^\dagger\mathsf U-2I$~\cite{cvetkovic_line,mielke_line}. The classification here identifies that familiar incidence structure within the compatible matching-layer Hamiltonian and includes the entire strength-dependent dark sector.

\subsection{An exact dark-sector flag from heterogeneous even-cycle decorations}
\label{sec:cycle_chain_flag}
We now construct an exactly growing dark sector within the hypotheses of Theorem~\ref{thm:general_matching_dark}.  Let $\mathscr G_n$ consist of $n$ vertex-disjoint even cycles $C_{\ell_k}$, $\ell_k\ge4$, joined in a chain by $n-1$ single bridges; on each intermediate cycle, the two bridge endpoints are different vertices.  Colour every cycle alternately with colours $1,2$ and every bridge with colour $3$.  This is a proper edge colouring with $q=3$, independently of $n$ and the cycle lengths.  The graphs are connected bipartite cacti, with
\begin{equation}
 |V(\mathscr G_n)|=\sum_{k=1}^n\ell_k,\qquad
 |E(\mathscr G_n)|=\sum_{k=1}^n\ell_k+n-1,\qquad \beta_1(\mathscr G_n)=n.
 \label{eq:decoration_cycle_counts}
\end{equation}

\begin{proposition}[Complete growing port-dark flag]
\label{prop:decoration_dark_flag}
Equip each $\mathscr G_n$ with the active-edge-end Hamiltonian and active-channel-symmetric ports of Theorem~\ref{thm:general_matching_dark}, with $J_e>0$ and $C_v=aI_{d_v}+b_v\mathbf1_v\mathbf1_v^\dagger$.  Assume that the strengths and conversion parameters on existing edges and vertices are interpreted consistently between stages, except for the required enlargement of the converter block at attachment vertices, and that the real scalar $a$ is common across stages.  Suppose no constituent cycle has all its kinetic strengths equal.  Then the complete port-dark space satisfies
\begin{equation}
 \mathscr D_n=\operatorname{span}\{\psi_1,\ldots,\psi_n\},\qquad
 \dim\mathscr D_n=n,\qquad H_n|_{\mathscr D_n}=aI,
 \label{eq:decoration_dark_flag}
\end{equation}
where the normalized $\psi_k$ has alternating symmetric-edge-end amplitudes on $C_{\ell_k}$ and vanishes on all other edges.  Zero extension of edge-end amplitudes gives a strict flag $\iota_n\mathscr D_n\subsetneq\mathscr D_{n+1}$ and
\begin{equation}
 H_{n+1}\iota_n\psi=\iota_nH_n\psi\quad(\psi\in\mathscr D_n).
 \label{eq:decoration_dark_intertwining}
\end{equation}
The assertion remains valid for different cycle lengths, nonuniform positive edge strengths, and independent real $b_v$.
\end{proposition}
\textbf{Proof.} In each even cycle choose edge coefficients $z_e=+1,-1,+1,-1,\ldots$ successively, zero on every other edge.  Their endpoint sums vanish at every vertex, including bridge endpoints.  The corresponding symmetric-edge state is therefore in $\iota_+(\ker\mathsf U)$, and all $n$ such states have disjoint supports.  By~(\ref{eq:decoration_cycle_counts}) and~(\ref{eq:general_unsigned_rank}), $\dim\ker\mathsf U=n$, so they span the symmetric dark sector.  Each bridge is a cut edge and the only simple cycles in $\mathscr G_n$ are the constituent motif cycles.  If no motif cycle has constant strength, then every equal-strength subgraph $F_J$ is a forest, and $\beta_1(F_J)=0$.  The complete classification~(\ref{eq:general_dark_count}) yields~(\ref{eq:decoration_dark_flag}).  Under zero extension, each old cycle state still has symmetric endpoint amplitudes, zero vertex sums, and no new-edge amplitudes; hence $H_{n+1}\iota_n\psi=a\iota_n\psi=\iota_nH_n\psi$.  The new cycle contributes one orthogonal vector, proving strictness. \hfill Q.E.D.

\begin{proposition}[Exact protected relative-phase control]
\label{prop:protected_relative_phase_control}
Let $\mathscr G_n$ and $\mathscr D_n=\operatorname{span}\{\psi_1,\ldots,\psi_n\}$
be as in Proposition~\ref{prop:decoration_dark_flag}. Let $\Pi_k$ be the
orthogonal projector onto all active edge-end states belonging to the cycle
$C_{\ell_k}$. For arbitrary locally integrable real functions $\lambda_k(t)$,
introduce the cycle-selective diagonal control
\begin{equation}
 V(t)=\sum_{k=1}^n\lambda_k(t)\Pi_k,
 \qquad H_{\rm ctrl}(t)=H(t)+V(t).
 \label{eq:protected_cycle_phase_control}
\end{equation}
Assume that $H(t)$ has arbitrary positive time-dependent kinetic strengths
and converters of the form~(\ref{eq:general_converter_family}) with common
real $a(t)$. Then $\mathscr D_n$ is invariant under $H_{\rm ctrl}(t)$, and
its exact propagator is
\begin{equation}
 U_{\mathscr D_n}(t,0)=\sum_{k=1}^n
 \exp\!\left[-\frac{i}{\hbar}\int_0^t\bigl(a(s)+\lambda_k(s)\bigr)\,ds\right]
 |\psi_k\rangle\langle\psi_k|.
 \label{eq:protected_cycle_phase_propagator}
\end{equation}
In particular, all $n-1$ independent relative phases can be prescribed
without leakage from $\mathscr D_n$.
\end{proposition}
\textbf{Proof.} Every $\psi_k$ has symmetric edge-end amplitudes and
vanishing port sum, hence $H(t)\psi_k=a(t)\psi_k$ for every allowed kinetic
strength. Disjoint cycle supports give $\Pi_j\psi_k=\delta_{jk}\psi_k$.
Thus $H_{\rm ctrl}(t)\psi_k=(a(t)+\lambda_k(t))\psi_k$ at all times.
The restricted generators share a fixed diagonal basis, so direct integration
gives~(\ref{eq:protected_cycle_phase_propagator}). Choosing the time
integrals of $\lambda_k-\lambda_1$ independently specifies the relative
phases. \hfill Q.E.D.

\noindent\textbf{Local-control restriction.} The disjoint cycle supports
constrain protected mixing. Every vertex-local Hermitian control
$W=\bigoplus_v W_v$ satisfies
$\langle\psi_l|W|\psi_k\rangle=0$ for $k\ne l$; if it leaves
$\mathscr D_n$ invariant, its restriction is diagonal. The conclusion
continues to hold for controls also coupling channel states across
individual bridge endpoints, provided $W\mathscr D_n\subseteq\mathscr D_n$.
Indeed, a bridge term acting on $\psi_k$ has support in the neighbouring
cycle only at its attachment vertex. If $W\psi_k$ lies in
$\mathscr D_n$, its component on that neighbouring cycle must be
proportional to $\psi_l$, which has nonzero amplitudes throughout the
cycle. The proportionality coefficient therefore vanishes.
Thus exact protected mixing requires controls outside this locality
class or a protocol with a varying protected subspace.

The complementary sector retains bridge-mediated dynamics. Let $uv$ be a cycle edge of colour $\alpha\in\{1,2\}$ incident at a bridge endpoint $v$, and let the bridge connect $v$ to $w$. If $b_v\ne0$, then
\begin{equation}
 \langle w,3|H_n^3|u,\alpha\rangle
 =J_{uv}\,b_v\,J_{vw}\ne0.
 \label{eq:decoration_active_bridge}
\end{equation}
The three-step transition consists of kinetic transfer along $uv$, channel
conversion at $v$, and transfer along $vw$. Thus protected cycle states
coexist with active inter-cycle transport. In block form, exact zero-extension
intertwining for a protected vector $\psi$ requires the interface operator
to satisfy $T\psi=0$, rather than $T=0$ on the full old Hilbert space.

For a countably infinite chain with uniformly bounded $J_e$, $|b_v|$,
and common $|a|$, the bounded-degree active-edge-end Hamiltonian defines a
bounded self-adjoint operator. The finitely supported $\psi_k$ remain
independent eigenvectors with eigenvalue $a$, giving infinite multiplicity.

The decorated family therefore has a protected subspace whose dimension
grows by one with each added cycle while the number of internal channel labels
remains fixed at three. These states are exactly invariant under arbitrary
time-dependent kinetic-strength variations, and cycle-selective diagonal
controls implement independent relative phases without leakage. The
complementary sector simultaneously supports bridge-mediated dynamics.
Compact cycle-localized states have established precedents in line-graph
and flat-band constructions~\cite{mielke_line,leykam_flat}; the feature here
is their explicit realization and controlled invariance within the
compatible matching-layer system.

\paragraph{The nonplanar $K_{3,3}$ specialization.}
The nonplanar graph $K_{3,3}$ provides an explicit converter-rigidity classification.  Because $K_{3,3}$ is $3$-regular and properly $3$-edge-colourable, its active subspace is the full $18$-dimensional physical Hilbert space.  The general theorem guarantees four symmetric cycle-dark modes; the next result characterizes all Hermitian onsite converters that preserve the complete four-dimensional space.

Let $G=K_{3,3}$ with vertex parts $\mathcal V_A=\{A_1,A_2,A_3\}$ and $\mathcal V_B=\{B_1,B_2,B_3\}$, and colour the edge $\{A_i,B_j\}$ by $c(A_i,B_j)=1+((j-i)\bmod 3)$.  The three colours form perfect matchings.  Assign weights $w_{A_iB_j}=\gamma_{ij}\varepsilon_{c(A_i,B_j)}$, $\gamma_{ij}>0$.  In the channel-resolved Hilbert space $\mathbb C^6\otimes\mathbb C^3$, the edge-end amplitudes of a state may be written as two matrices $(X,Y)\in\mathbb C^{3\times3}\oplus\mathbb C^{3\times3}$, where $X_{ij}$ and $Y_{ij}$ correspond to the $A_i$ and $B_j$ ends of edge $A_iB_j$ in its selected channel.  With $J_{ij}=\hbar^2\gamma_{ij}^2/m>0$, the squared-Laplacian kinetic Hamiltonian acts by
\begin{equation}
 (H_0(X,Y))_{X,ij}=J_{ij}(X_{ij}-Y_{ij}),\qquad
 (H_0(X,Y))_{Y,ij}=J_{ij}(Y_{ij}-X_{ij}).
 \label{eq:k33_edge_kinetic}
\end{equation}
Let $\mathsf B_{\rm port}:\mathbb C^{18}\to\mathbb C^6$ denote the unnormalised channel-summation map,
\begin{equation}
 (\mathsf B_{\rm port}(X,Y))_{A_i}=\sum_jX_{ij},\qquad
 (\mathsf B_{\rm port}(X,Y))_{B_j}=\sum_iY_{ij}.
 \label{eq:k33_port_map}
\end{equation}
Thus the normalized channel-symmetric vertex-port map is $\mathsf P_{\rm port}^\dagger=\mathsf B_{\rm port}/\sqrt3$.  We call a subspace port-dark when it is Hamiltonian-invariant and annihilated by $\mathsf B_{\rm port}$.  This is a notion relative to these ports; a channel-resolved measurement need not be dark.  Define
\begin{equation}
 \mathcal C=\{Z\in\mathbb C^{3\times3}:\sum_j Z_{ij}=0,\ \sum_i Z_{ij}=0\},\qquad
 \mathscr D_{\rm cyc}=\{(Z,Z):Z\in\mathcal C\}.
 \label{eq:k33_cycle_space}
\end{equation}
The five independent row/column constraints give $\dim\mathcal C=4$, the cycle rank $9-6+1$ of $K_{3,3}$.  The identification uses its bipartite incidence structure.

\begin{theorem}[Rigidity of the $K_{3,3}$ cycle-dark sector]
\label{thm:k33_rigidity}
For arbitrary $J_{ij}>0$, let $H=H_0+\bigoplus_{v\in V(G)}C_v$, with each $C_v=C_v^\dagger\in M_3(\mathbb C)$ acting at one vertex on its three channel amplitudes.  The entire four-dimensional subspace $\mathscr D_{\rm cyc}$ is invariant if and only if there are real $a$ and $b_v$ such that
\begin{equation}
 C_v=aI_3+b_v\mathbf1\mathbf1^\dagger\quad (v\in V(G)),
 \qquad \mathbf1=(1,1,1)^T.
 \label{eq:k33_converter_class}
\end{equation}
In that case $H\psi=a\psi$ and $\mathsf B_{\rm port}\psi=0$ for every $\psi\in\mathscr D_{\rm cyc}$, independently of the nine kinetic strengths and six parameters $b_v$.  If the converter matrices are required to have zero diagonal, this condition reduces to the common off-diagonal converter $C_v=g(\mathbf1\mathbf1^\dagger-I_3)$, with $a=-g$ and the same $g$ at all vertices.
\end{theorem}
\textbf{Proof.} Every symmetric edge-end state $(Z,Z)$ is annihilated by $H_0$ edge by edge.  If~(\ref{eq:k33_converter_class}) holds, the three amplitudes at each vertex of a cycle state sum to zero, so $C_v$ acts on them as $aI_3$; this proves sufficiency and darkness.
For necessity, fix one edge $A_iB_j$.  There exists a four-cycle amplitude $Z\in\mathcal C$ with its $j$-th column zero, but its $i$-th row a nonzero vector supported on the other two columns with opposite entries.  Invariance of $(Z,Z)$ under the onsite operator requires the transformed amplitude at the $A_i$ end of the chosen edge to vanish, since the input amplitude at its $B_j$ end is zero.  Varying $j$ shows that, on the local zero-sum plane $\mathbf1^\perp$, each output coordinate of $C_{A_i}$ depends only on the corresponding input coordinate.  Because invariance also preserves zero-sum vectors, these three coordinate multipliers are equal; hence $C_{A_i}|_{\mathbf1^\perp}=a_{A_i}I$.  The argument at the $B_j$ vertices is identical.  Every edge lies in a four-cycle with nonzero amplitude on that edge, so equality of its two transformed edge-end amplitudes gives $a_{A_i}=a_{B_j}$.  Connectivity forces a common real $a$.  Hermiticity then makes $\mathbf1$ an invariant orthogonal complement of $\mathbf1^\perp$, proving~(\ref{eq:k33_converter_class}).  Zero diagonal imposes $a+b_v=0$ at every vertex.\hfill Q.E.D.

\begin{theorem}[Complete port-dark multiplicity in the protected converter family]
\label{thm:k33_complete_dark}
For each kinetic value $J$, let $F_J$ have edge set $E(F_J)=\{e:J_e=J\}$. Under~(\ref{eq:k33_converter_class}), the maximal port-dark invariant subspace is
\begin{equation}
 \mathscr D_E(H,\mathsf B_{\rm port})=\mathscr D_{\rm cyc}\oplus\bigoplus_J
 \big\{(Z,-Z):\operatorname{supp}Z\subseteq E(F_J),\ \mathsf B_{\rm port}(Z,-Z)=0\big\},
 \label{eq:k33_dark_decomp}
\end{equation}
and its dimension is
\begin{equation}
 \dim\mathscr D_E(H,\mathsf B_{\rm port})=4+\sum_J\beta_1(F_J).
 \label{eq:k33_dark_dimension}
\end{equation}
Here $\beta_1(F_J)=|E(F_J)|-|V(F_J)|+c(F_J)$ is the cycle rank.  The first summand has energy $a$, and the $J$-summand has energy $2J+a$.  In particular, distinct edge strengths yield exactly four dark states; nine equal strengths yield eight.
\end{theorem}
\textbf{Proof.} Equation~(\ref{eq:k33_converter_class}) gives $C=aI+\mathsf B_{\rm port}^\dagger\operatorname{diag}(b_v)\mathsf B_{\rm port}$, so $C=aI$ on $\ker\mathsf B_{\rm port}$.  Every port-dark invariant subspace is therefore an $H_0$-invariant subspace of $\ker\mathsf B_{\rm port}$, and conversely.  Since $H_0$ is Hermitian, its maximal such subspace is a direct sum of intersections of $\ker\mathsf B_{\rm port}$ with its eigenspaces.  Each matching edge contributes one symmetric kinetic eigenvector of eigenvalue $0$ and one antisymmetric eigenvector of eigenvalue $2J_{ij}$.  In the symmetric sector, $\mathsf B_{\rm port}=0$ gives the four-dimensional space~(\ref{eq:k33_cycle_space}).  In the antisymmetric sector at eigenvalue $2J$, $\mathsf B_{\rm port}=0$ is the oriented incidence-kernel condition on $F_J$; because $F_J$ is bipartite, the kernel has dimension $\beta_1(F_J)$.  Adding the scalar $a$ proves the spectrum and the asserted completeness.\hfill Q.E.D.

The theorem permits heterogeneous off-diagonal conversion when compensated by onsite potentials: the choice $C_v=g_v(\mathbf1\mathbf1^\dagger-I_3)+\mu_v I_3$ preserves all four modes precisely when $\mu_v-g_v$ is independent of $v$.  For time-dependent positive edge strengths and converters $C_v(t)=a(t)I_3+b_v(t)\mathbf1\mathbf1^\dagger$, every cycle-dark state undergoes only the common phase $\exp[-(i/\hbar)\int a(s)\,ds]$.  The complete converter characterization and multiplicity formula identify
which local couplings preserve the four structural dark modes and when
additional equal-strength modes occur. The incidence-kernel mechanism is
well established in spectral graph theory and flat-band constructions
~\cite{cvetkovic_line,hatsugai_flatbands,leykam_flat}; the inverse
converter characterization is specific to the compatible matching-layer
Hamiltonian.

\begin{ack}
I acknowledge Mihály Bányai for useful discussions at the very beginning of this project and I am grateful to my colleagues András Telcs, Péter Érdi, Zoltán Somogyvári, Tamás Kiss, László Zalányi, Marcell Stippinger and Attila Bencze for convincing me to demonstrate the applicability of the approach.
\end{ack}


\begin{thebibliography}{99}
\bibitem{ab_positive} C. D. Aliprantis and O. Burkinshaw, \textit{Positive Operators}, Springer, Dordrecht, 2006. \url{https://doi.org/10.1007/978-1-4020-5008-4}
\bibitem{anderson_ctmc} W. J. Anderson, \textit{Continuous-Time Markov Chains: An Applications-Oriented Approach}, Springer, New York, 1991. \url{https://doi.org/10.1007/978-1-4612-3038-0}
\bibitem{arveson} W. Arveson, \textit{An Invitation to C*-Algebras}, Graduate Texts in Mathematics \textbf{39}, Springer, New York, 1976. \url{https://doi.org/10.1007/978-1-4612-6371-5}
\bibitem{barenco} A. Barenco, C. H. Bennett, R. Cleve, D. P. DiVincenzo, N. Margolus, P. Shor, T. Sleator, J. A. Smolin and H. Weinfurter, \textit{Elementary gates for quantum computation}, Phys. Rev. A \textbf{52} (1995), 3457--3467. \url{https://doi.org/10.1103/PhysRevA.52.3457}
\bibitem{bla} F. Bazsó and E. Lábos, \textit{Boolean--Lie algebras and the Leibniz rule}. J. Phys. A: Math. Gen. \textbf{39}(2006), 6871--6876. \url{http://dx.doi.org/10.1088/0305-4470/39/22/005}
\bibitem{benichou} O. B\'enichou, C. Loverdo, M. Moreau and R. Voituriez, \textit{Intermittent search strategies}, Rev. Mod. Phys. \textbf{83} (2011), 81--129. \url{https://doi.org/10.1103/RevModPhys.83.81}
\bibitem{berkuch} G. Berkolaiko and P. Kuchment, \textit{Introduction to Quantum Graphs}, MSM \textbf{186}, American Mathematical Society, Providence, Rhode Island, 2013. \url{https://doi.org/10.1090/surv/186}
\bibitem{biggs_graph} N. Biggs, \textit{Algebraic Graph Theory}, 2nd ed., Cambridge University Press, Cambridge, 1993.
\bibitem{birkhoff_pierce} G. Birkhoff and R. S. Pierce, \textit{Lattice-ordered rings}, Anais da Academia Brasileira de Ciencias \textbf{28} (1956), 41--69.
\bibitem{bose} S. Bose, \textit{Quantum communication through an unmodulated spin chain}, Phys. Rev. Lett. \textbf{91} (2003), 207901. \url{https://doi.org/10.1103/PhysRevLett.91.207901}
\bibitem{christandl} M. Christandl, N. Datta, A. Ekert and A. J. Landahl, \textit{Perfect state transfer in quantum spin networks}, Phys. Rev. Lett. \textbf{92} (2004), 187902. \url{https://doi.org/10.1103/PhysRevLett.92.187902}
\bibitem{ch} F. Chung, \textit{Laplacians and the Cheeger inequality for directed graphs}. Ann. Comb. \textbf{9}(2005), 1--19. \url{http://dx.doi.org/10.1007/s00026-005-0237-z}
\bibitem{cvetkovic_line} D. Cvetkovi\'{c}, P. Rowlinson and S. Simi\'{c}, \textit{Spectral Generalizations of Line Graphs: On Graphs with Least Eigenvalue $-2$}, Cambridge University Press, Cambridge, 2004. \url{https://doi.org/10.1017/CBO9780511751752}
\bibitem{debruijn_erdos} N. G. de Bruijn and P. Erd\H{o}s, \textit{A colour problem for infinite graphs and a problem in the theory of relations}, Proc. Koninkl. Nederl. Akad. Wetensch. Ser. A \textbf{54} (1951), 371--373.
\bibitem{diestel} R. Diestel, \textit{Graph Theory}, 5th ed., Graduate Texts in Mathematics \textbf{173}, Springer, Berlin, 2017. \url{https://doi.org/10.1007/978-3-662-53622-3}
\bibitem{dim} A. Dimakis and F. Müller--Hoissen, \textit{Discrete differential calculus: Graphs, topologies, and gauge theory}. J. Math. Phys. \textbf{35}(1994), 6703--6735. \url{http://dx.doi.org/10.1063/1.530638}
\bibitem{farinati} M.A. Farinati and A.P. Jancsa, \textit{Lie bialgebra structures on 2-step nilpotent graph algebras}. J. of Alg. \textbf{505}(2018), 70-91. \url{http://dx.doi.org/10.1016/j.jalgebra.2018.03.003}
\bibitem{gj} L. Gillman and M. Jerison, \textit{Rings of Continuous Functions}, Graduate Texts in Mathematics \textbf{43}, Springer, New York, 1976 [originally published by Van Nostrand, 1960]. \url{https://doi.org/10.1007/978-1-4615-7819-2}
\bibitem{grebenkov2019} D. S. Grebenkov, \textit{A unifying approach to first-passage time distributions in diffusing diffusivity and switching diffusion models}, J. Phys. A: Math. Theor. \textbf{52} (2019), 174001. \url{https://doi.org/10.1088/1751-8121/ab0dae}
\bibitem{hatsugai_flatbands} Y. Hatsugai, \textit{Revisiting flat bands and localization}, Ann. Phys. \textbf{435} (2021), 168453. \url{https://doi.org/10.1016/j.aop.2021.168453}
\bibitem{hein_graphstates} M. Hein, J. Eisert and H. J. Briegel, \textit{Multiparty entanglement in graph states}, Phys. Rev. A \textbf{69} (2004), 062311. \url{https://doi.org/10.1103/PhysRevA.69.062311}
\bibitem{kadison_ringrose} R. V. Kadison and J. R. Ringrose, \textit{Fundamentals of the Theory of Operator Algebras, Vol. I: Elementary Theory}, Academic Press, New York, 1983.
\bibitem{kauf} L.H. Kauffman, \textit{Non--Commutative Worlds}. New J. Phys. \textbf{6}(2004), 173. \url{http://dx.doi.org/10.1088/1367-2630/6/1/173}
\bibitem{yks} Y. Kosmann-Schwarzbach, \textit{Lie bialgebras, Poisson Lie groups and dressing transformations}.  In: Y. Kosmann-Schwarzbach, B. Grammaticos and K.M. Tamizhmani (Eds.), \textit{Integrability of Nonlinear Systems}, LNP \textbf{638}, Springer-Verlag Berlin Heidelberg 2004, pp. 107--173. \url{http://dx.doi.org/10.1007/978-3-540-40962-5_5}
\bibitem{leykam_flat} D. Leykam, A. Andreanov and S. Flach, \textit{Artificial flat band systems: from lattice models to experiments}, Adv. Phys. X \textbf{3} (2018), 1473052. \url{https://doi.org/10.1080/23746149.2018.1473052}
\bibitem{zaanen} W. A. J. Luxemburg and A. C. Zaanen, \textit{Riesz Spaces I}, North-Holland, Amsterdam, 1971; A. C. Zaanen, \textit{Riesz Spaces II}, North-Holland, Amsterdam, 1983.
\bibitem{maas_entropy} J. Maas, \textit{Gradient flows of the entropy for finite Markov chains}, J. Funct. Anal. \textbf{261} (2011), 2250--2292. \url{https://doi.org/10.1016/j.jfa.2011.06.009}
\bibitem{majid} S. Majid, \textit{Noncommutative Riemannian geometry on graphs}. J. of Geom. and Phys. \textbf{69}(2013), 74-93. \url{http://dx.doi.org/10.1016/j.geomphys.2013.02.004}
\bibitem{mielke_line} A. Mielke, \textit{Ferromagnetism in the Hubbard model on line graphs and further considerations}, J. Phys. A: Math. Gen. \textbf{24} (1991), 3311--3321. \url{https://doi.org/10.1088/0305-4470/24/14/018}
\bibitem{nagumo} M. Nagumo, \textit{Über die Lage der Integralkurven gewöhnlicher Differentialgleichungen}, Proc. of the Physico-Mathematical Soc. of Japan. 3rd Series, \textbf{24}(1942), 551-559.  \url{https://doi.org/10.11429/ppmsj1919.24.0_551}
\bibitem{nov} S.P. Novikov, \textit{Schrödinger operators on graphs and symplectic geometry}. Fields Institute Commun. \textbf{24}(1999), 397–413. \url{http://dx.doi.org/10.1090/fic/024/23}
\bibitem{redner} S. Redner, \textit{A Guide to First-Passage Processes}, Cambridge University Press, Cambridge, 2001. \url{https://doi.org/10.1017/CBO9780511606014}
\bibitem{rovelli2019} C. Rovelli and V. Zatloukal, \textit{Natural discrete differential calculus in physics}. Found. of Phys. \textbf{49}(2019), 693–699. \url{http://dx.doi.org/10.1007/s10701-019-00271-1}
\bibitem{sun} T. Sunada, \textit{Discrete geometric analysis}. In: P. Exner, J. P. Keating, P. Kuchment, T. Sunada and A. Teplyaev, (Eds.), \textit{Analysis on graphs and its applications} Proc. of Symposia in Pure Mathematics, \textbf{77}, 2008, pp. 51--83.  \url{http://dx.doi.org/10.1090/pspum/077/2459864}
\bibitem{vidal} J. Vidal, R. Mosseri and B. Dou\c{c}ot, \textit{Aharonov--Bohm cages in two-dimensional structures}, Phys. Rev. Lett. \textbf{81} (1998), 5888--5891. \url{https://doi.org/10.1103/PhysRevLett.81.5888}
\end{thebibliography}
\end{document}